\documentclass[11pt]{article}
\pdfoutput=1
\usepackage{jheppub}
\usepackage{tabu}
\usepackage[vcentermath]{youngtab}
\usepackage[usenames,dvipsnames,table]{xcolor}
\usepackage{graphicx,amsmath,amssymb,amsthm,multirow,array,bm,bbm,esint}
\usepackage[mathscr]{eucal}
\usepackage[bbgreekl]{mathbbol}
\usepackage{epsf,amsfonts}
\usepackage{slashed}
\usepackage[numbers,sort&compress]{natbib}
\usepackage{minibox}
\usepackage{rotating}
\usepackage{pdflscape}
\usepackage{array,tikz-cd}
\usepackage{subcaption}
\usepackage{framed}
\usepackage{mathtools}
\usepackage{tikz} 
\usepackage[compat=1.1.0]{tikz-feynman}
\usepackage{feynmf}
\usepackage{float}
\usepackage{verbatim}

\usepackage{slashed}

\newcommand{\Phib}{	\overline{\Phi} }
\newcommand{\etaflc}{ \eta }

\def\ben{\begin{equation}}
\def\een{\end{equation}}

\def\res{\mathop{\text{Res}}\limits}        

\newcommand{\cD}{{\mathcal{G}}}     

\newcommand{\biL}{{\Box_{\rm bi}}}      

\newcommand{\cpw}{ \psi }    

\newcommand{\norm}{ \mathcal{N} }     

\newcommand{\cdim}{ c }     

\newcommand{\normalization}{ \Upsilon_{\mathcal{O}} }     

\newcommand{\dop}{\widehat{\mathfrak{D}}_s}        

\newcommand{\thomas}{\mathsf{D} }       

\newcommand{\kdif}{ k_{\text{\tiny dif}} }
\newcommand{\kdifb}{ \bar{k}_{\text{\tiny dif}} }

\def\JY#1{{\color{RubineRed}{\bf #1}}}

\newcommand{\ie}{\emph{i.e.}, }

\newcommand{\etc}{\emph{etc}.}
\newcommand{\vs}{\emph{vice versa}}

\makeatletter
\newcommand{\doubletilde}[1]{{%
  \mathpalette\double@tilde{#1}%
}}
\newcommand{\double@tilde}[2]{%
  \sbox\z@{$\m@th#1\tilde{#2}$}%
  \ht\z@=.9\ht\z@
  \tilde{\box\z@}%
}
\makeatother

\title{AdS Maps and Diagrams of Bi-local Holography}

\author[a,b]{Robert de Mello Koch}
\author[c]{\!, Antal Jevicki}
\author[c,d]{\!, Kenta Suzuki}
\author[e]{\!, Junggi Yoon}
\affiliation[\,a]{School of Physics and Telecommunication Engineering, South China Normal University, Guangzhou 510006, China.}
\affiliation[\,b]{National Institute for Theoretical Physics, School of Physics and Centre for Theoretical Physics,\\ 
University of the Witwatersrand, Wits, 2050, South Africa.}
\affiliation[\,c]{Department of Physics, Brown University, 182 Hope Street, Providence, RI 02912, U.S.A.}
\affiliation[\,d]{Centre de Physique Th\'{e}orique, \'{E}cole Polytechnique, CNRS, F-91128 Palaiseau, France.}
\affiliation[\,e]{School of Physics, Korea Institute for Advanced Study\\
85 Hoegiro Dongdaemun-gu, Seoul 02455, Republic of Korea.}
\emailAdd{robert@neo.phys.wits.ac.za}
\emailAdd{antal\_jevicki@brown.edu}
\emailAdd{kenta.suzuki@polytechnique.edu}
\emailAdd{junggiyoon@kias.re.kr}

\preprint{{\raggedleft \tt BROWN-HET-1769\par } \par\par {\raggedleft \tt KIAS-P18094 \par}  }

\abstract{We present in detail the basic ingredients contained in bi-local holography, representing a constructive scheme for reconstructing AdS bulk theories in Vectorial/AdS duality. Explicit Mapping to bulk AdS and higher spin fields is seen to be given by a double Fourier transform. All order interactions are explicitly specified through the collective action. This generates bulk Feynman (Witten) diagrams (at tree and loop level). We give details of the four-point case evaluation. It is noted that the bi-local construction goes beyond the
assumptions in various discussions of non-locality.
}

\begin{document}
\maketitle

\section{Introduction}
\label{sec:intro}

The approach to Holography based on collective fields represents one constructive scheme to understand the appearance of emergent space-time degrees of freedom and of specifying their bulk interactions~\cite{Das:2003vw, Koch:2010cy, Koch:2014aqa, Koch:2014mxa}.
Collective (gauge invariant) fields with their intrinsic non-locality and the feature of remaining closed under large $N$ Schwinger-Dyson (SD) equations provide a natural framework for bulk reconstruction.
Vector type theories, leading to AdS Higher Spin (HS) Duality~\cite{Klebanov:2002ja, Sezgin:2002rt, Giombi:2009wh, Maldacena:2011jn} are interesting and possibly the simplest class of theories where this construction can be explicitly implemented.
The space of collective fields is in this case given by bi-locals whose nonlinear dynamics
and equations of motion are completely specified by the SD equations generating the $1/N$ expansion.
This is summarized by the corresponding collective action (and Hamiltonian). It was proposed and in subsequent studies established that the bi-local space contains precisely the sufficient degrees of freedom to allow a representation of bulk fields and their dynamics in AdS space-time~\cite{Das:2003vw, Koch:2010cy, Koch:2014aqa, Koch:2014mxa}.
The notable feature is that the identification of (physical) bulk fields is one-to-one and given by a momentum-space map through a double Fourier transform.
This was explicitly constructed in the canonical (Hamiltonian) framework in both light-cone~\cite{Koch:2010cy} and time-like~\cite{Koch:2014aqa} quantization. Interesting features dealing with $1/N$ were discussed at both tree and loop level~\cite{Giombi:2013fka,Jevicki:2014mfa}.

This framework, for reconstruction of bulk space-time and field interactions based on the dynamics of bi-local fields is termed bi-local holography. At present, it represents the only one-to-one construction for understanding the emergence of AdS degrees of freedom and construction of their bulk interactions from large $N$ CFT. With the previous studies~\cite{Koch:2010cy,Koch:2014aqa} being in the canonical (single time) framework (of collective theory), it is the purpose of this work to establish the more general, covariant action formulation and map.
In the lower, one-dimensional case ($d=1$) this has recently been extensively used in the investigations and large $N$ solution of SYK type models~\cite{Polchinski:2016xgd,Jevicki:2016bwu, Maldacena:2016hyu,Das:2017wae}. While the complete bulk gravitational theory is not yet established, the bi-local construction and map greatly illuminate the emergent space-time and degrees of freedom of this theory.

So in the present work we consider the higher dimensional cases of vector models in $d>1$ at both UV and IR critical points (with clear dualities to Higher Spin Gravities in AdS~\cite{Fronsdal:1978rb, Fang:1978wz,Fronsdal:1978vb,Vasiliev:1990en, Vasiliev:1995dn,Bekaert:2005vh}) in more detail. For this we give
a Map that establishes a transformation to AdS space-time, while the nonlinear collective action generates interaction vertices to all orders. Their structure is based on a bi-local star product, with an
extended dipole picture~\cite{Jevicki:2011ss} analogous to the case of SFT~\cite{Gross:1986ia}. As such the bi-local theory provides a nonlinear bulk description of Higher Spins to all orders.
This allows construction and evaluation of Feynman (Witten) diagrams which we will exemplify with detailed consideration of general 4-point boundary amplitudes.
The content of the paper goes as follows.
We will first discuss the covariant Map from bi-locals to AdS HS fields.
This Map will be constructed in general in Section~\ref{sec:bi-local map to ads} and in detail for the example of AdS$_3$.
It follows the earlier momentum-space Fourier analysis and applies to any case of vectorial/AdS HS duality.
This will also be true for the structure of bi-local interactions and the associated Feynman diagrams.
We will discuss this in Section~\ref{sec:bi-local witten diagrams} presenting the details of the 4-point case.
It is interesting to compare this with the work and evaluations of~\cite{Bekaert:2015tva} and questions of Non-locality~\cite{Sleight:2017pcz,Ponomarev:2017qab}.

\section{Overview of Bi-local Holography}
\label{sec:overview of bi-local holography}

Bi-local holography provides a scheme of constructing various higher spin type bulk AdS theories from large $N$ vector type CFT's.
It is based on the collective field approach to the $1/N$ expansion where one exactly represents the path integral
    \begin{equation}
        Z \, = \, \int D\phi_i\, e^{-S[\phi]} \, ,
    \end{equation}
in terms of a new bi-local set of invariant variables $\Phi$
    \begin{equation}
        Z \, = \, \int D\Phi\, \mu (\Phi)\, e^{-NS_{\rm col}[\Phi]} \, .
    \end{equation}
Here we have rescaled $\Phi$ to exhibit an overall $N$ which then plays the role of a coupling constant. In this way, the bi-local method systematically describes the large $N$ physics (in vector type models) by an action~\cite{Das:2003vw}
	\begin{equation}
		S_{\rm col}[\Phi] \, = \, S[\Phi] \, - \, \frac{1}{2} \, {\rm Tr} \log\Phi \, , 
	\label{S_col}
	\end{equation}
where the trace term comes from a Jacobian factor due to the change of path-integral variables from the fundamental fields $\phi_i$ ($i=1, \cdots, N$) to the bi-local collective field
	\begin{equation}
		\Phi(x_1, x_2) \, = \, \frac{1}{N} \sum_{i=1}^N \phi_i(x_1) \phi_i(x_2) \, .
	\end{equation}
The first term $S[\Phi]$\footnote{For the $d$-dimensional $O(N)$ vector model, we have
	\begin{equation}
		S[\Phi] \, = \, \frac{1}{2} \, \int d^dx \bigg[- \Big(\nabla_x^2 \Phi(x, x') \Big)_{x'=x} + \frac{\lambda}{2} \Phi^2(x, x) \bigg] \, .
	\end{equation} } represents the original action of the vector model written in terms of $\Phi$.
The saddle-point solution $\Phi_0$ of the collective action~\eqref{S_col} represents the two-point function of the fundamental fields
	\begin{equation}
		\Phi_0(x_1, x_2) \, = \, \frac{1}{N} \sum_{i=1}^N \big\langle \phi_i(x_1) \phi_i(x_2) \big\rangle \, .
	\end{equation}
Expanding the collective action around the saddle-point solution as follows
	\begin{equation}
		\Phi(x_1, x_2) \, = \, \Phi_0(x_1, x_2) \, + \, |x_{12}|^2{1\over \sqrt{N}} \ \Phib(x_1, x_2) \, ,
	\label{fluctuations}
	\end{equation}
with $x_{12}\equiv x_1 - x_2$, we find the quadratic action
	\begin{equation}
		S_{(2)} \, = \, \frac{1}{2} \, \int \prod_{k=1}^4 d^dx_k \ \Phib(x_1, x_2) \, \widehat{\mathcal{L}}_{\rm bi} \, \Phib(x_3, x_4) \, .
	\end{equation}
The bi-local Laplacian is given by
	\begin{equation}
		\biL \, = \, |x_{12}|^2 |x_{34}|^2 
		\Big[ \Phi_0^{-1}(x_1, x_3) \Phi_0^{-1}(x_2, x_4) \, + \, {\lambda\over 2} \, \delta^d(x_{12}) \delta^d(x_{13}) \delta^d(x_{14}) \Big] \, .
	\end{equation}
We will be following the critical theories at $\lambda_{\rm UV}=0$, $\lambda_{\rm IR}$. The bi-local generator of the conformal group is given by
\begin{equation}
    L_{AB}=L_{AB}^1+L_{AB}^2\, .
\end{equation}
For the Euclidean $d$-dimensional conformal field theory,
the two Casimir operators of the $SO(1,d+1)$ conformal symmetry, whose generators are denoted by $L_{AB}$ ($A,B = -1, 0 , \cdots, d$), are defined by 
	\begin{align}
		C_2 \, \equiv& \, \frac{1}{2} L_{AB} L^{AB} \, ,\\
		C_4 \, \equiv& \, \frac{1}{2} \widetilde{C}_4 \, - \, \frac{1}{2} C_2^2 
	\end{align}
with
	\begin{equation}
		\widetilde{C}_4 \, = \, \frac{1}{2} L_A{}^B L_B{}^C L_C{}^D L_D{}^A \, .
	\end{equation}
The eigenvalues of these operators acting on a primary with conformal dimension $\Delta$ and spin $s$ are given in \cite{Dolan:2011dv}\footnote{Note that we use Hermitian $SO(1,d+1)$ generators. Hence, the eigenvalue of the quadratic Casimir $C_2$ has opposite sign with respect to that in~\cite{Dolan:2011dv}.}. The bi-local Laplacian (for the free theory) is related to the Casimir operators as \cite{Koch:2014mxa}
	\begin{equation}
		\biL \,\equiv \,{1\over 4} |x_{12}|^4 \, \nabla_{x_1}^2\nabla_{x_2}^2\,=  \, C_4 + {1\over 4} C_2^2 + {d^2-3d+4\over 4} C_2 \, .\label{def: bilocal laplacian}
	\end{equation}
The second equality can be shown by explicit evaluation. One considers the eigenvalue problem of the bi-local Laplacian
\begin{equation}
		\biL \, \cpw_{\cdim, s}(x_1, x_2) \, = \, \lambda_{\cdim, s} \, \cpw_{\cdim,s}(x_1, x_2) \, .
	\label{eigen-eq}
\end{equation}
where we label the eigenfunction and its eigenvalue by $\cdim$ and $s$, with the dimension $\Delta$ related to $c$ via $\Delta\equiv{d\over 2}+\cdim$. The eigenfunction $\cpw_{\cdim, s}$, which is called a conformal partial wave function~\cite{Todorov}, is given by a particular set of conformal spinning three point functions
\begin{equation}
	\cpw_{ \cdim,s} (x_1,\cdim_0;x_2,\cdim_0;x_3,\cdim,s,\varepsilon)\sim  { (Z_{12,3}\cdot  \varepsilon )^s\over |x_{12}|^{2\Delta_0 - \Delta} |x_{23}|^{\Delta} |x_{31}|^\Delta }\label{eq: intro cpw}
\end{equation}
where $\varepsilon^\mu$ denotes a null polarization vector and $Z^\mu_{12,3}$ is a unit vector defined by
\begin{equation}
	Z^\mu_{12,3} \equiv {|x_{13}||x_{23}|\over |x_{12}| }\left(  {x_{13}^\mu \over |x_{13}|^2}- {x_{23}^\mu \over |x_{23}|^2} \right)\, .
\end{equation}
Also note that $\Delta_0\equiv {d\over 2} +\cdim_0 \equiv\frac{d-2}{2}$ is the conformal dimension of the $d$-dimensional free scalar field. The eigenvalues of the bi-local Laplacian $\biL$ are given by 
\begin{equation}
		\lambda_{\cdim, s} \, = \, {1\over 4}\left[ \left(\Delta-{d\over 2}\right)^2 - \left(s+{d\over 2}-2\right)^2\right]\left[ \left(\Delta-{d\over 2}\right)^2 - \left(s+{d\over 2}\right)^2 \right]\, .
	\label{eigenvalues}
\end{equation}
We discuss the properties of the conformal partial wave functions $\cpw_{\cdim,s} (x_1,\cdim_0;x_2,\cdim_0;x_3,\cdim,s,\varepsilon)$ in Appendix~\ref{app: conformal partial wave}.

The main feature of the bi-local reconstruction of the bulk AdS fields is that there exists a map between $\etaflc$ $(\equiv\, |x_{12}|^2\, \Phib)$
and the bulk higher spin fields $H$
	\begin{align}
		\etaflc(x_1^i, x_2^i) \, &= \, \int_{\rm AdS} \mathcal{M}(x_1^i, x_2^i| x^i, z, S) \, H(x^i, z, S) \, , \nonumber\\
		H(x^i, z, S) \, &= \, \int_{\rm Bi-local} \mathcal{M}^{-1}(x^i, z, S| x_1^i, x_2^i) \, \etaflc(x_1^i, x_2^i) \, ,
	\end{align}
where $x^i$ and $z$ are the AdS coordinates and $S$ denotes the spin degrees of freedom.
This bulk AdS field $H({\rm AdS}_{d+1},S_{d-1})$ contains physical HS fields whose meaning will be illuminated in Section \ref{sec:bi-local map to ads}.
We therefore have a one-to-one identification of the AdS$_{d+1}$ bulk from the bi-local field.
Under the transformation, the collective Laplacian turns into its corresponding AdS form and one can consider the collective action as an action of bulk AdS higher spin theory.

The sequence of interacting  vertices follows from this action.
One expands the action around the saddle-point as in~\eqref{fluctuations}, which leads to the $1/N$ expansion of the collective action as
	\begin{align}
		S[\eta] \, &= \, \frac{1}{2} \, \int d^dx_1 d^dx_2 d^dx_3 d^dx_4 \, \Phib(x_1, x_2) \mathcal{K}(x_1, x_2; x_3, x_4) \Phib(x_3, x_4) \nonumber\\
		&\qquad + \, \sum_{n=3}^{\infty} {1\over N^{{n\over 2}-1}} {(-1)^n \over 2 n } \,
		{\rm Tr}\Big(\Phi_0^{-1} \star \etaflc \star \cdots \star \Phi_0^{-1} \star \etaflc \Big) \, ,
    \label{S-expansion}
	\end{align}
where in the interaction term (second term), the $n^{\rm \emph{\tiny th}}$ term in the sum contains $n$ $\etaflc$ fields.
Recall that $\etaflc(x_1, x_2) \equiv |x_{12}|^2 \, \Phib(x_1, x_2)$. The star product is defined by $(A \star B)(x_1, x_2) \equiv \int d^dx_3 \, A(x_1, x_3) B(x_3, x_2)$.
The quadratic kernel is the  bi-local Laplacian $\mathcal{K}=\biL$, given by
equation (\ref{def: bilocal laplacian}).

On the other hand, vertices are constructed using the inverse of the classical solution $\Phi_0^{-1}$, which are represented by a dashed line. For example, the cubic and quartic vertices are diagrammatically presented by 
\begin{gather}
    \mathcal{V}_3(x_1,x_2;x_3,x_4;x_5,x_6)\;\equiv\; \begin{minipage}[c]{0.2\linewidth}
\centering
    \begin{tikzpicture}
        \draw [line width=1.2pt, dashed,domain=-15:75] plot ({0.7*cos(\x)}, {0.7*sin(\x)});
        \draw [line width=1.2pt, dashed,domain=105:195] plot ({0.7*cos(\x)}, {0.7*sin(\x)});
        \draw [line width=1.2pt, dashed,domain=-45:-135] plot ({0.7*cos(\x)}, {0.7*sin(\x)});
        \draw [fill] ({0.7*cos(75)}, {0.7*sin(75)}) circle [radius=.08];
        \draw [fill] ({0.7*cos(-15)}, {0.7*sin(-15)}) circle [radius=.08];
        \draw [fill] ({0.7*cos(105)}, {0.7*sin(105)}) circle [radius=.08];
        \draw [fill] ({0.7*cos(195)}, {0.7*sin(195)}) circle [radius=.08];
        \draw [fill] ({0.7*cos(-45)}, {0.7*sin(-45)}) circle [radius=.08];
        \draw [fill] ({0.7*cos(-135)}, {0.7*sin(-135)}) circle [radius=.08];
        \draw (0,0) node {\textit{3}};
        \draw ({0.7*cos(105)-0.3}, {0.7*sin(105)+0.2}) node {$x_1$};
        \draw ({0.7*cos(75)+0.3}, {0.7*sin(75)+0.2}) node {$x_2$};
        \draw ({0.7*cos(-15)+0.3}, {0.7*sin(-15)+0.1}) node {$x_3$};
        \draw ({0.7*cos(-45)+0.3}, {0.7*sin(-45)-0.1}) node {$x_4$};
        \draw ({0.7*cos(-135)-0.3}, {0.7*sin(-135)-0.1}) node {$x_5$};
        \draw ({0.7*cos(195)-0.3}, {0.7*sin(195)+0.1}) node {$x_6$};
\end{tikzpicture} 
\end{minipage}\label{eq: cubic vertex diagram}\\
    \mathcal{V}_4(x_1,x_2;x_3,x_4;x_5,x_6;x_7,x_8)\;\equiv\; \begin{minipage}[c]{0.2\linewidth}
\centering
    \begin{tikzpicture}
        \draw [line width=1.2pt, dashed,domain=15:75] plot ({0.7*cos(\x)}, {0.7*sin(\x)});
        \draw [line width=1.2pt, dashed,domain=105:165] plot ({0.7*cos(\x)}, {0.7*sin(\x)});
        \draw [line width=1.2pt, dashed,domain=195:255] plot ({0.7*cos(\x)}, {0.7*sin(\x)});
        \draw [line width=1.2pt, dashed,domain=-15:-75] plot ({0.7*cos(\x)}, {0.7*sin(\x)});
        \draw [fill] ({0.7*cos(15)}, {0.7*sin(15)}) circle [radius=.08];
        \draw [fill] ({0.7*cos(75)}, {0.7*sin(75)}) circle [radius=.08];
        \draw [fill] ({0.7*cos(105)}, {0.7*sin(105)}) circle [radius=.08];
        \draw [fill] ({0.7*cos(165)}, {0.7*sin(165)}) circle [radius=.08];
        \draw [fill] ({0.7*cos(195)}, {0.7*sin(195)}) circle [radius=.08];
        \draw [fill] ({0.7*cos(255)}, {0.7*sin(255)}) circle [radius=.08];
        \draw [fill] ({0.7*cos(-15)}, {0.7*sin(-15)}) circle [radius=.08];
        \draw [fill] ({0.7*cos(-75)}, {0.7*sin(-75)}) circle [radius=.08];
        \draw (0,0) node {\textit{4}};
        \draw ({0.7*cos(105)-0.2}, {0.7*sin(105)+0.2}) node {$x_1$};
        \draw ({0.7*cos(75)+0.2}, {0.7*sin(75)+0.2}) node {$x_2$};
        \draw ({0.7*cos(15)+0.3}, {0.7*sin(15)+0.1}) node {$x_3$};
        \draw ({0.7*cos(-15)+0.3}, {0.7*sin(-15)-0.1}) node {$x_4$};
        \draw ({0.7*cos(-75)+0.2}, {0.7*sin(-75)-0.2}) node {$x_5$};
        \draw ({0.7*cos(255)-0.2}, {0.7*sin(255)-0.2}) node {$x_6$};
        \draw ({0.7*cos(195)-0.3}, {0.7*sin(195)-0.1}) node {$x_7$};
        \draw ({0.7*cos(165)-0.3}, {0.7*sin(165)+0.1}) node {$x_8$};
\end{tikzpicture} 
\end{minipage}\label{eq: quartic vertex diagram}
\end{gather}  
We will  work out concrete forms of these vertices and their transforms to AdS space to exemplify their locality properties. They will also be the main ingredients for writing and evaluating Witten diagrams in Section~\ref{sec:bi-local witten diagrams}. For the $d=1$ case relevant to the SYK model these vertices generate the diagrammatics of~\cite{Gross:2017aos}.

\section{Bi-local Map to AdS}
\label{sec:bi-local map to ads}

The basis of the map comes from group theory and concerns the action
of the conformal group. On the bi-local $\Phi(x_1,x_2)$ we have the kinematic action of two copies of $SO(1,d+1)$
\begin{equation}
    L_{AB} \, = \, L_{AB}^1+L_{AB}^2 \, ,
\end{equation}
and on the higher spin fields $H({\rm AdS}_{d+1},S_{d-1})$ there is a nontrivial realization (spinning particle in AdS) of $SO(1,d+1)$
\begin{equation}
    {\cal M}=\sum_{\Delta,s,p^\mu}\psi_{\rm AdS} ({\rm AdS}_{d+1},S_{d-1}|\Delta,s,p_\mu)\psi_{\rm bi}(x_1,x_2|\Delta,s,p_\mu) \, .
\end{equation}
Under the bi-local map, the mapping between the bi-local field and the AdS higher spin field
is simply given by a change of momenta (point transformation) in momentum space.
Namely, defining the Fourier transform to momentum space by 
	\begin{align}
		\widetilde{\etaflc}(k_1^i, k_2^i)
		\, &\equiv \, \int dx_1^i dx_2^i \, e^{ix_1 \cdot k_1} e^{ix_2 \cdot k_2} \, \etaflc(x_1^i, x_2^i) \, , \nonumber\\
		\widetilde{H}(p^i, p^z, \Theta) \, &\equiv \, \int dx^i dz dS \, e^{ip\cdot x} e^{ip^z z} e^{i \Theta S} \, H(x^i, z, S) \, ,
	\end{align}
the bi-local map transforms between the two fields
	\begin{equation}
	    \widetilde{\etaflc}(k_1^i, k_2^i) \quad \Longleftrightarrow \quad \widetilde{H}(p^i, p^z, \Theta) \, .
	\end{equation}
To understand this identification, we first need to establish the physical degrees of freedom of the bulk AdS theory .

Higher spin fields $h_{\mu_1\cdots \mu_s}(x)$ ($\mu_j=0,1, \cdots, d$) in AdS$_{d+1}$ can be represented in embedding space by
	\begin{equation}
		H_{A_1\cdots A_s}(X) \, = \, \sum_s {x^{\mu_1}}_{, A_1}\cdots {x^{\mu_1}}_{,A_s} h_{\mu_1\cdots \mu_s}(x) \, , 
	\end{equation}
where $X^A$ ($A=-1,0,1,\cdots, d$) are the coordinates of the embedding space of AdS$_{d+1}$. Introducing an extra set of coordinates $Y^A$ to generate the spin degrees of freedom, we have
	\begin{equation}
		H(X;Y) \, \equiv \, \sum_{s} H_{A_1\cdots A_s}(X)\; Y^{A_1}\cdots Y^{A_s} \, .
	\end{equation}
The phase space is spanned by
	\begin{equation}
		X^A\;,\;\; P_A \;,\;\; Y^A\;,\;\; K_A \, ,
	\end{equation}
(where $P_A$ and $K_A$ are the conjugate momentum of $X^A$ and $Y^A$, respectively). On this space one  has four (first class) constraints which can be summarized as~\cite{Fronsdal:1978vb,Koch:2014mxa}
	\begin{alignat}{3}
		&T_1 \, = \, &&X^AP_A +Y^A K_A+1 \, = \, 0 \, , \qquad&& \mbox{(massless condition)} \\
		&T_2 \, = \, &&X^AK_A \, = \, 0 \, , &&\mbox{(transversality condition)}\\
		&T_3 \, = \, &&K^A K_A \, = \, 0 \, , &&\mbox{(traceless condition)}\\
		&T_4 \, = \, &&P^AK_A \, = \, 0 \, , &&\mbox{(de Donder gauge)}
	\end{alignat}
together with two ``gauge conditions''
	\begin{alignat}{3}
		&T_{-1} \, = \, &&X^A X_A +1 \, = \, 0 \, , \hspace{10mm}&&\mbox{(AdS Space)}\\
		&T_{-2} \, = \, &&X^A Y_A \, = \, 0 \, . \hspace{10mm} &&
	\end{alignat}
By solving the constraints, the phase space is reduced to $(x^\mu,p_\mu;y^\mu,k_\mu)$ ($\mu=0,1,\cdots, d$). In the reduced phase space, the remaining first class constraints can be written as
	\begin{align}
		T_3 \, &= \, \delta_{\mu\nu} k^\mu k^\nu z^2\, , \\ 
		T_4 \, &= \, z^2\delta_{\mu\nu} k^\mu p^\nu + z(\delta_{\mu\nu}k^\mu k^\nu)y^2 \, .
	\end{align}
Note that $T_3$ and $T_4$ can be considered as constraints for the higher spin field $\phi(x;y)$ in the Poincare coordinates of AdS$_{d+1}$ space
	\begin{equation}
		\phi(x;y) \, \equiv \, \sum_s \phi_{\mu_1\cdots \mu_s}(x)\; y^{\mu_1}\cdots y^{\mu_s} \, .
	\end{equation}
Solving $T_3$ and $T_4$ leads to a reduced phase space for AdS$_{d+1}\times S^{d-1}$ with coordinates $(x^\mu,p_\mu;\Omega,p^\Omega)$. This is the physical space, to which we will refer as the spinning AdS space. On this phase space there is no gauge redundancy, and the action of the group $SO(1,d+1)$ is nontrivial.

Our correspondence corresponds to this (physical) description of AdS higher spin bulk fields.
The transformation between bi-local and spinning AdS spaces is simplified with the Poincare parametrization of AdS.
One has that the relationship between bi-local and spinning AdS spaces is simply given as a change of co-ordinates in momentum space. This change of momenta, and the explicit form of the map can be determined through comparison of the actions of the conformal group on the two spaces: bi-local and Spinning AdS. We will present a demonstration of this in the following section.

\subsection{Momentum Map}
\label{sec:momentum map}

Let us give the discussion for deducing the momentum map in the example of the $d=2$ and AdS$_3$ $\times$ $S^1$ case. By solving the remaining constraints $T_3$ and $T_4$~\cite{Koch:2014mxa}, the realization of $SO(1,3)$ on  the reduced space can be established to be
	\begin{align}
		L_+^{\rm AdS} \, &= \, ip \, , \nonumber\\
		L_0^{\rm AdS} \, &= \, ipx + {i\over 2}p^z z \, , \nonumber\\
		L_-^{\rm AdS} \, &= \, ipx^2 + i x z p^z - i z^2 \bar{p} + i{ (p^\theta)^2\over 4p} + {\sqrt{4p \bar{p} +(p^z)^2}\over 2 p}  z p^\theta  \, , \nonumber\\
		\bar{L}_+^{\rm AdS} \, &= \, i\bar{p} \, ,\nonumber\\
		\bar{L}_0^{\rm AdS} \, &= \, i\bar{p} \bar{x} + {i\over 2}p^z z \, , \nonumber\\
		\bar{L}_-^{\rm AdS} \, &= \, i\bar{p} \bar{x}^2 + i\bar{x} z p^z - i z^2 p +  i{ (p^\theta)^2\over 4\bar{p}}   - {\sqrt{4p \bar{p} + (p^z)^2}\over 2 \bar{p} } zp^\theta \label{eq: ads gens} \, ,
	\end{align}
where $p \equiv {1\over 2}(p^0 -i p^1)$ and $\bar{p} \equiv {1\over 2}(p^0 + i p^1)$. Note that $\theta$ and $p^\theta$ corresponds to the coordinate for $S^1$ and its conjugate, respectively.
%
For the bi-local space, the $SO(1,3)$ generators read
	\begin{align}
		L_+^{\rm bi} \, &= \, ik_1 + ik_2 \, , \nonumber\\
		L_0^{\rm bi} \, &= \,- k_1 \frac{\partial}{\partial k_1} \, - \, k_2 \frac{\partial}{\partial k_2} \, , \nonumber\\
		L_-^{\rm bi} \, &= \, - i k_1 \left( \frac{\partial}{\partial k_1} \right)^2 \, - \, ik_2 \left( \frac{\partial}{\partial k_2} \right)^2 \, .
	\end{align}
For the bi-local CFT, we use the notation $k_i\equiv {1\over 2} (k_i^0-i k_i^1)$, $\bar{k}_i\equiv {1\over 2} (k_i^0+i k_i^1)$ and $\kdif\equiv k_1-k_2$, $\kdifb\equiv \bar{k}_1-\bar{k}_2$.
One can search for the momentum space map by requiring that the bi-local realization (of generators) as differential operator maps into the AdS$_3$ $\times$ $S^1$ space realization.
This has been done in the previous light-cone and canonical quantizations for AdS$_4$, and the covariant $d=1$ example of AdS$_2$.
Generally this is an exceedingly complicated exercise. We will describe a much simpler route, which will allow us to deduce the momentum space transformation relatively easily through comparison of momentum space wave functions. One then proceeds to verify the consistency of the transformation by comparison of all the symmetry generators.

For the derivation of the bi-local map, it is actually convenient to consider the Lorentzian signature case. Therefore, until \eqref{eq: inv momentum map-Lorentzian} (and also Appendix~\ref{app:fourier transform of spinning correlators}), we consider Lorentzian signature while in the rest of the paper we focus on the Euclidean case. This Lorentzian version of the bi-local map~\eqref{eq: lorentzian momentum map1}$\sim$\eqref{eq: lorentzian momentum map3} can be analytically continued to the Euclidean version~\eqref{eq: euclidean momentum map1}$\sim$\eqref{eq: euclidean momentum map3}. In Appendix~\ref{app:checks for the bi-local map}, we also confirm that this bi-local map for the Euclidean case transforms the $SO(1,3)$ generators for AdS into bi-local generators.

Let us consider the Casimir operators for Lorentzian $SO(2,2)$ generators in Appendix~\ref{app: realization of so(2,2)}
\begin{equation}
		C_2 \, = \, 2(L^2 + \overline{L}^2) \, , \qquad  \, C_4 \, = \, - (L^2 - \overline{L}^2)^2 
\end{equation}
with
\begin{equation}
		L^2 \, = \, -\frac{1}{4} \Big( z \sqrt{4 p_+ p_- - p_z^2} + p_{\theta} \Big)^2 \, , \qquad
		\overline{L}^2 \, = \, -\frac{1}{4} \Big( z \sqrt{4 p_+ p_- - p_z^2} - p_{\theta} \Big)^2 \, .
\end{equation}
Here we define $p_\pm = -{1\over 2}(p^0\mp p^1)$ for Lorentzian case. The Laplacian squared is given by\footnote{In this section, we neglect the ordering term which is linear in $C_2$. This corresponds to take $\Delta, s \gg1$ for the eigenvalues \eqref{eigenvalues} and we have
	\begin{equation}
		\lambda_{\Delta, s} \, \approx \, \frac{1}{4} (\Delta^2-s^2)^2 \, .
	\end{equation}}
	\begin{equation}
		\Box_{{\rm AdS}_3 \times S^1}^2 \, = \, C_4 + \frac{1}{4} \, C_2^2 \, = \, 4 L^2 \overline{L}^2 \, .
	\end{equation}
To see that this formula makes sense, recall that from CFT we expect
	\begin{equation}
		L^2 \, = \,- \frac{1}{4} \, (\Delta+s)^2 \, , \qquad \overline{L}^2 \, = \, -\frac{1}{4} \, (\Delta-s)^2 \, , 
	\end{equation}
so that the eigenvalues of the Laplacian squared are $[{1\over 2}(-\Delta^2+s^2)]^2$ as expected. Using the explicit forms given above, one finds the Laplacian as
	\begin{equation}
		\Box_{{\rm AdS}_3 \times S^1} \, = \,- \frac{1}{2} \Big[ \big( z \sqrt{ 4 p_+ p_- - p_z^2} \, \big)^2 - \, p_{\theta}^2 \Big] \, .
	\end{equation}
By changing $z \to i\frac{\partial}{\partial p_z}$ and $p_{\theta} \to -i\frac{\partial}{\partial \theta}$, we have
 	\begin{equation}
		\Box_{{\rm AdS}_3 \times S^1} \, = \, \frac{1}{2} \left[ \frac{\partial^2}{\partial \phi^2} \, - \, \frac{\partial^2}{\partial \theta^2} \right] \, ,
	\end{equation}
where we define
 	\begin{equation}
		\phi \, \equiv \, - \arcsin \left( \frac{p_z}{\sqrt{4p_+p_-}} \right) \, .\label{def: lorentzian phi}
	\end{equation}
Then, it is easy to see that eigenfunctions 
 \begin{equation}
		\psi_{\Delta, s} \, = \, e^{\pm  (i \Delta \phi +is\theta)} \label{eq: Lorentzian ads wave function}
\end{equation}
lead to the eigenvalues $(-\Delta^2+s^2)/2$ as expected.

For the bi-local (Lorentzian) CFT, the bi-local forms of the Casimir operators are given by
 	\begin{equation}
		L^2 \, = \, \left( \frac{\partial}{\partial k_{1+}} - \frac{\partial}{\partial k_{2+}} \right)^2
		k_1^+k_2^+\, , \qquad
		\overline{L}^2 \, = \,  \left( \frac{\partial}{\partial k_{1-}} - \frac{\partial}{\partial k_{2 -} } \right)^2
	   k_{1-}k_{2-} \, ,
	\end{equation}
where we use the notation $k_{i\pm} \equiv -\frac{1}{2}(k_i^0 \mp k_i^1)$ for the Lorentzian case. The Laplacian $\biL=4L^2\bar{L}^2$ becomes
 	\begin{equation}
		\Box_{\rm bi} \, = \, 2 \left( \frac{\partial}{\partial k_{1+}} - \frac{\partial}{\partial k_{2+} } \right) \left( \frac{\partial}{\partial k_{1-} } - \frac{\partial}{\partial k_{2-} } \right)\sqrt{k_{1+} k_{2+} k_{1-} k_{2-} } \, , 
	\end{equation}
From the Fourier transformation of the spinning three-point functions~\eqref{3pt in momentum}, we have the bi-local wave functions
 	\begin{equation}
		\psi(\vec{k}_1, \vec{k}_2) \, = \, \delta(\vec{k}_1+\vec{k}_2-\vec{p}) F({\kdif}_+) F({\kdif}_-) \, ,
	\end{equation}
where ${\kdif}_\pm \equiv k_{1\pm} - k_{2\pm}$ and 
 	\begin{equation}
		F(\kdif^+) F(\kdif^-) \, = \, \frac{e^{i \frac{s}{2}(\arcsin\frac{{\kdif}_+}{p_+}-\arcsin\frac{{\kdif}_-}{p_-})} \, e^{i \frac{\Delta}{2}(\arcsin\frac{{\kdif}_+}{p_+}+\arcsin\frac{{\kdif}_-}{p_-})}}
		{\sqrt{k_{1+}k_{2+}k_{1-}k_{2-} } }\, .
	\end{equation}
Therefore, one can confirm that $\psi(\vec{k}_1, \vec{k}_2)$ is indeed the bi-local wave function with the expected eigenvalue.
 	\begin{equation}
		\Box_{\rm bi} \, F({\kdif}_+) F({\kdif}_-) \, = \,  \frac{1}{2} (-\Delta^2 + s^2) F({\kdif}_+) F({\kdif}_-) \, .
	\end{equation}
By comparing with the AdS wave function in \eqref{eq: Lorentzian ads wave function}, we conclude that
 	\begin{align}
		\theta \, =& \, {1\over 2}\arcsin \frac{{\kdif}_+}{p_+} \, - {1\over 2} \, \arcsin \frac{{\kdif}_-}{p_-} \, ,\\
		\phi \, =& \, {1\over 2}\arcsin \frac{{\kdif}_+}{p_+} \, + \,{1\over 2} \arcsin \frac{{\kdif}_-}{p_-} \, .
	\end{align}

In order to write down the map for $p_z$, it is useful to define
	\begin{equation}
		\tan\alpha_+ \equiv \, \sqrt{\frac{k_{2+}}{k_{1+}}} \quad , \qquad \tan\alpha_- \equiv \, \sqrt{\frac{k_{2-}}{k_{1-}}} \ ,
	\end{equation}
Note that we have the following identity:
	\begin{equation}
		\tan\left( \arcsin{{\kdif}_\pm\over p_\pm} \right) \, = \, \frac{{\kdif}_\pm}{2\sqrt{k_{1\pm} k_{2\pm}}} \, = \, \cot 2\alpha_\pm \, .
	\end{equation}
This leads to
	\begin{equation}
		\arcsin{{\kdif}_\pm\over p_\pm} \, = \, \frac{\pi}{2} \, - \, 2\alpha_\pm \, + \, \pi n_\pm \, ,
	\end{equation}
where $n_\pm\in \mathbb{Z}$. Choosing $n_\pm=0$, the map for $\phi$ gives
	\begin{equation}
		p^z \, = \, - 2\sqrt{p_+ p_-} \sin\phi \, = \, - 2\sqrt{p_+ p_-}\cos(\alpha_++\alpha_-) \, = \, - 2\sqrt{k_{1+} } \sqrt{k_{1-} } + 2\sqrt{k_{2+} } \sqrt{k_{2-} } \, .
	\end{equation}
In summary, we have the Lorentzian version of the momentum map from $(k_{1\pm};k_{2\pm} )$ to $(p_\pm,p^z,\theta)$
	\begin{align}
		p_\pm \, &= \, k_{1\pm} + k_{2\pm} \, , \label{eq: lorentzian momentum map1}\\
		p^z \, &= \, - 2\sqrt{k_{1+} } \sqrt{k_{1-} } + 2\sqrt{k_{2+} } \sqrt{k_{2-} } \, , \label{eq: lorentzian momentum map2}\\
		\theta \, &= \, - \arctan {\sqrt{k_{2+} }\over \sqrt{k_{1+} }} + \arctan {\sqrt{k_{2-} }\over\sqrt{ k_{1-} } } \, , \label{eq: lorentzian momentum map3}
	\end{align}
and its inverse map is given by
	\begin{equation}
		k_{1\pm} \, = \, \frac{p_\pm}{2} \Big[ 1 + \sin( \phi\pm\theta) \Big] \quad , \qquad k_{2\pm} \, = \, \frac{p_\pm}{2} \Big[ 1 - \sin( \phi\pm\theta) \Big] \, .\label{eq: inv momentum map-Lorentzian}
	\end{equation}
For the Euclidean bi-local map, we perform a Wick rotation to Euclidean signature
\begin{equation}
    k_+\;,\; k_-\;,\;p_+\;,\; p_- \quad\longrightarrow \quad e^{-{i\pi \over 2}}k\;,\; e^{-{i\pi \over 2}}\bar{k}\;, \; e^{-{i\pi \over 2}}p \;,\; e^{-{i\pi \over 2}}\bar{p} \, ,\label{eq: wick rotation}
\end{equation}
and quickly obtain the Euclidean version of the momentum map from $(k_1, \bar{k}_1; k_2, \bar{k}_2)$ to $(p,\bar{p},p^z,\theta)$
	\begin{align}
		p \, &= \, k_1 + k_2 \, , \label{eq: euclidean momentum map1}\\
		\bar{p} \, &= \, \bar{k}_1 + \bar{k}_2 \, , \label{eq: euclidean momentum map1bar}\\
		p^z \, &= \, 2i \sqrt{k_1} \sqrt{\bar{k}_1} - 2i\sqrt{k_2} \sqrt{\bar{k}_2} \, , \label{eq: euclidean momentum map2}\\
		\theta \, &= \, -\arctan{\sqrt{k_2}\over \sqrt{k_1}} \, + \, \arctan {\sqrt{\bar{k}_2}\over \sqrt{\bar{k}_1} } \, , \label{eq: euclidean momentum map3}
	\end{align}
and its inverse map, given by
\begin{alignat}{3}
	k_1\, =&\, {p\over 2}\left[1+\sin (\phi +\theta )\right]\quad,\qquad k_2\, =&\, {p\over 2}\left[1-\sin (\phi +\theta )\right]\ ,\label{eq: inv momentum map1}\\
	\bar{k}_1\, =&\, {\bar{p}\over 2}\left[1+\sin (\phi -\theta )\right]\quad,\qquad \bar{k}_2\, =&\, {\bar{p}\over 2}\left[1-\sin (\phi -\theta )\right]\, .\label{eq: inv momentum map2}
\end{alignat}
By Wick rotation, one can also repeat the analogous derivation for the Euclidean bi-local map. From the $SO(1,3)$ generators for higher spin fields in \eqref{eq: ads gens}, the Casimir operators are written as
\begin{equation}
		L^2 \, = \, \frac{1}{4} \Big( z \sqrt{p_z^2+4p\bar{p}} + i   p_{\theta} \Big)^2 \, , \qquad \overline{L}^2  \, = \, \frac{1}{4} \Big( z \sqrt{p_z^2+4p\bar{p}} -i  p_{\theta} \Big)^2 \, .
\end{equation}
This leads to the Laplacian squared
\begin{equation}
	\Box_{{\rm AdS}_3 \times S^1}^2 \, =  \, 4 L^2 \overline{L}^2 \,=\, \left[ {1\over 2}\big(z\, \sqrt{\, p_z^2 +4p\bar{p}} \, \big)^2\, + \, {1\over 2}p_{\theta}^2\right]^2 \, = \, {1\over 2}\left(\frac{\partial^2}{\partial \phi^2} \, - \,  \frac{\partial^2}{\partial \theta^2}\right) \, .
\end{equation}
where $\phi$ is defined by Wick rotation in \eqref{eq: wick rotation} of \eqref{def: lorentzian phi}
\begin{equation}
	i\phi \, = \, \sinh^{-1} \left( \frac{p_z}{\sqrt{4p \bar{p} }} \right) \, .
\end{equation}
One can express the Casimir operators in terms of $\phi$ and $\theta$
\begin{equation}
		L^2 \, = \, \frac{1}{4} \Big( \partial_\phi + \partial_{\theta}   \Big)^2 \, , \qquad \overline{L}^2 \, = \, \frac{1}{4} \Big(  \partial_\phi - \partial_\theta \Big)^2 \, ,
\end{equation}
and, it is easy to see that $\cpw_{\Delta, s} =  e^{i\Delta \phi+is\theta  }$ is the eigenfunction of $L^2$ and $\overline{L}^2$ corresponding to eigenvalue $- \frac{1}{4} \, (\Delta+s)^2$ and $- \frac{1}{4} \, (\Delta-s)^2$, respectively. As in the Lorentzian case, one can express $\theta$ and $\phi$ in terms of $\kdif,\kdifb, p$ and  $\bar{p}$ by comparing the wave functions.
\begin{align}
	\theta \, =& \, {1\over 2}\arcsin \frac{\kdif}{p} \, - \,{1\over 2} \arcsin \frac{\kdifb}{\bar{p}} \, ,\\
	\phi \, =&  {1
		\over 2}\, \arcsin \frac{\kdif}{p} \, + \, {1\over 2}\arcsin \frac{\kdifb}{\bar{p}} \, .
\end{align}	
This also leads to the Euclidean bi-local map in~\eqref{eq: euclidean momentum map1}$\sim$\eqref{eq: euclidean momentum map3}. As usual in Euclidean CFT$_2$, holomorphic and anti-holomorphic momenta can be considered as independent complex variables. In the physical space where they are conjugate to each other, the bi-local map in \eqref{eq: euclidean momentum map2} and \eqref{eq: euclidean momentum map3} implies that $p^z$ and $\theta$ (and their conjugates in \eqref{eq: coordinate map3} and \eqref{eq: coordinate map4}) are pure imaginary.

Note that the bi-local map is not necessarily a canonical transformation in general. Nevertheless, demanding that the two phase spaces are related by a canonical transformation, this momentum map induces a coordinate map, which is explicitly given in Appendix~\ref{app:checks for the bi-local map}. Then, the bi-local map consisting of the momentum and coordinate map indeed transforms $SO(1,3)$ generators for higher spin field to bi-local ones and \vs.

In fact, one can also derive the bi-local map for momentum and coordinates by equating the two realization of $SO(1,3)$ generators. The bi-local map which we present in \eqref{eq: euclidean momentum map1}$\sim$\eqref{eq: euclidean momentum map3} and \eqref{eq: coordinate map1}$\sim$\eqref{eq: coordinate map4} is a solution of these equations of the generators.

Through the bi-local map found above, the quadratic collective action for $d=2$ is mapped to AdS$_3$ as
	\begin{equation}
		S_{(2)} \, = \, \frac{1}{4} \, \int d^3x d\theta \, \sqrt{g} \,
		H \Big(\Box_{{\rm AdS}_3} + (\partial_{\theta} + i )^2 \Big) \Big(\Box_{{\rm AdS}_3} + (\partial_{\theta} - i )^2 \Big) H \, .
	\end{equation}
Hence, the AdS Laplacian reads
	\begin{equation}
		\widehat{\mathcal{L}}_{\rm AdS} \, = \, \Big(\Box_{\rm AdS} - m_1^2(s) \Big) \Big(\Box_{\rm AdS} - m_2^2(s) \Big) \, ,
	\end{equation}
where $m_1(s)$ represents the physical mass which corresponds to the scaling dimension $\Delta_1 = s+d-2$,
while $m_2(s)$ represents un-physical mass corresponding to $\Delta_2 = s+d$.

\subsection{Vertices in Momentum Space}
\label{sec:vertices in momentum space}

Start from the cubic term in the action
	\begin{equation}
		S_{(3)} \, = \, - \, \frac{1}{6\sqrt{N}} \, {\rm Tr}\Big(\Phi_0^{-1} \star \etaflc \star \Phi_0^{-1} \star \etaflc \star \Phi_0^{-1} \star \etaflc \Big) \, .
	\end{equation}
where we redefined the fluctuations as
	\begin{equation}
		\etaflc(x_1, x_1') \, = \, |x_1 - x_1'|^2 \, \Phib(x_1, x_1') \, .
	\label{redefinition}
	\end{equation}
Writing the interaction in terms of the $\Phib$ fields, one finds the cubic term is
	\begin{equation}
		S_{(3)}  =  -  \frac{1}{6\sqrt{N}} \, \prod_{i=1}^3 \Bigg[ \int d^2x_i d^2x_i' \, |x_i - x_i'|^2 \, \Phib(x_i, x_i') \Bigg] \,
		\Phi_0^{-1}(x_1', x_2) \Phi_0^{-1}(x_2', x_3) \Phi_0^{-1}(x_3', x_1) \, .
	\end{equation}
Now Fourier transforming the fluctuations 
	\begin{equation}
		\Phib(x_i, x_i') \, = \, \int \frac{d^2k_i}{(2\pi)^2} \frac{d^2k'_i}{(2\pi)^2} \ e^{ik_i x_i + i k'_i x'_i} \ \widetilde{\Phi}(k_i, k_i') \, ,
	\label{Fourier}
	\end{equation}
the cubic term reduces to
	\begin{align}
		&S_{(3)}  \cr
		=&  -  \frac{b^{-3}}{6\sqrt{N}}  \prod_{i=1}^3 \Bigg[ \int \frac{d^2k_id^2k_i'}{(2\pi)^4}  \, \sqrt{k_i^2 k_i'^2} \,
		\left| \frac{\partial}{\partial k_i} - \frac{\partial}{\partial k_i'} \right|^2 \, \widetilde{\Phi}(k_i, k_i') \Bigg] \, \delta^2(k_1'+k_2) \delta^2(k_2'+k_3) \delta^2(k_3'+k_1) \nonumber\\
		=& \, - \, \frac{b^{-3}}{6\sqrt{N}} \, \prod_{i=1}^3 \Bigg[ \int \frac{d^2k_id^2k_i'}{(2\pi)^4} \ \widetilde{\Phi}(k_i, k_i') \, \widehat{\mathcal{L}}^{\rm bi}_i(k_i, k_i') \Bigg] \,
		\delta^2(k_1'+k_2) \delta^2(k_2'+k_3) \delta^2(k_3'+k_1) \, ,
	\end{align}
where $\widehat{\mathcal{L}}^{\rm bi}_i$ is the bi-local Laplacian of the $i$-th leg acting on the delta functions.
Now mapping this cubic term to AdS$_3 \times S^1$, we obtain
	\begin{equation}
		S_{(3)}  =  -  \frac{b^{-3}}{6\sqrt{N}}  \prod_{i=1}^3 \Bigg[ \int \frac{dp^0_i dp^1_i dp^z_i d\theta_i}{(2\pi)^4} \ \mathcal{J}(\vec{p}_i, p^z_i, \theta_i) \, H(\vec{p}_i, p^z_i, \theta_i) \, 
		\widehat{\mathcal{L}}^{{\rm AdS}_3 \times S^1}_i(\vec{p}_i, p^z_i, \theta_i) \Bigg]   \delta^6(\cdots) \, ,
	\end{equation}
where the bi-local Laplacian is transformed into the AdS$_3 \times S^1$ Laplacian with a Jacobian $\mathcal{J}$.
Here, we identified $\widetilde{\Phi}(k_i, k_i') = H(\vec{p}_i, p^z_i, \theta_i)$ and $\delta^6(\cdots)$ represents the transformed results from
$\delta^2(k_1'+k_2) \delta^2(k_2'+k_3) \delta^2(k_3'+k_1)$.
Two of these delta functions can be written as $\delta^2(\vec{p}_1+\vec{p}_2+\vec{p}_3)$ and the remaining four are to be determined.

Let us now evaluate the $\delta^6(\cdots)$.
The best way to evaluate the transformation is to use the holomorphic and anti-holomorphic coordinates.
Then, the six delta functions are written as
	\begin{align}
		&\qquad \delta^2(k_1'+k_2) \delta^2(k_2'+k_3) \delta^2(k_3'+k_1) \\
		&= \, 2^{-6} \, \delta(k_1 + k_2') \delta(\bar{k}_1 + \bar{k}_2') \delta(k_2 + k_3') \delta(\bar{k}_2 + \bar{k}_3')
		\delta(k_3 + k_1') \delta(\bar{k}_3 + \bar{k}_1') \, . \nonumber
	\end{align}
Since the anti-holomorphic sector is just a copy of the holomorphic sector, we focus on the former
	\begin{align}
		({\rm holomorphic}) \, &\equiv \, 2^{-3} \, \delta(k_1 + k_2') \delta(k_2 + k_3') \delta(k_3 + k_1') \nonumber\\
		&= \, 2^{-1} \, \delta(k_1 + k_1' + k_2 + k_2' + k_3 + k_3') \nonumber\\
		&\quad \times \, \delta(k_1 + k_1' - k_2 + k_2' + k_3 - k_3') \nonumber\\
	    &\quad \times \, \delta(k_1 - k_1' + k_2 + k_2' - k_3 + k_3') \, . 
	\end{align}
Then, this part is transformed as
	\begin{align}
		({\rm holomorphic}) \, &\Rightarrow \, 
		2^{-1} \, \delta(p_1 + p_2 + p_3) \, \delta\Big(p_2 \big(1 + \sin(\phi_2 + \theta_2) \big) + p_3 \big(1 - \sin(\phi_3 + \theta_3) \big) \Big) \nonumber\\
		&\qquad \quad \times \, \delta\Big(p_1 \big(1- \sin(\phi_1 + \theta_1) \big) + p_3 \big(1+\sin(\phi_3 + \theta_3) \big) \Big) \, .
	\end{align}
We can obtain a similar result for the anti-holomorphic sector using the momentum bi-local map.

Now we further proceed to the quadratic coupling term
	\begin{equation}
		S_{(4)} \, = \, {1\over 8N} \, {\rm Tr}\Big(\Phi_0^{-1} \star \etaflc \star \Phi_0^{-1} \star \etaflc \star \Phi_0^{-1} \star \etaflc \star \Phi_0^{-1} \star \etaflc \Big) \, .
	\end{equation}
Again by redefining the fluctuation as~\eqref{redefinition} and Fourier transforming as~\eqref{Fourier}, we rewrite the quartic term as
	\begin{equation}
		S_{(4)} \, 
		= \, \frac{3 b^{-4}}{N} \, \prod_{i=1}^4 \Bigg[ \int \frac{d^2k_id^2k_i'}{(2\pi)^4}  \ \widetilde{\Phi}(k_i, k_i') \, \widehat{\mathcal{L}}^{\rm bi}_i(k_i, k_i') \Bigg] \,
		\delta^2(k_1'+k_2) \delta^2(k_2'+k_3) \delta^2(k_3'+k_4) \delta^2(k_4'+k_1) \, .
	\end{equation}
The transformation to AdS$_3\times S^1$ now leads to
	\begin{equation}
		S_{(4)} \, = \, \frac{ b^{-4}}{N} \, \prod_{i=1}^4 \Bigg[ \int \frac{dp^0_i dp^1_i dp^z_i d\theta_i}{(2\pi)^4} \ \mathcal{J}(\vec{p}_i, p^z_i, \theta_i) \, H(\vec{p}_i, p^z_i, \theta_i) \, 
		\widehat{\mathcal{L}}^{{\rm AdS}_3 \times S^1}_i(\vec{p}_i, p^z_i, \theta_i) \Bigg]  \, \delta^8(\cdots) \, .
	\end{equation}
The evaluation of $\delta^8(\cdots)$ is identical to the cubic case, and this gives
	\begin{align}
		({\rm holomorphic}) \, &\Rightarrow \,
		2^{-1} \, \delta(p_1 + p_2 + p_3 + p_4) \nonumber\\
		&\quad \times \, \delta\Big(p_2 \big(1+ \sin(\phi_2 + \theta_2) \big) + p_3 \big(1 - \sin(\phi_3 + \theta_3) \big) \Big) \nonumber\\
		&\quad \times \, \delta\Big(p_3 \big(1+ \sin(\phi_3 + \theta_3) \big) + p_4 \big(1 - \sin(\phi_4 + \theta_4) \big) \Big) \nonumber\\
		&\quad \times \, \delta\Big(p_4 \big(1+ \sin(\phi_4 + \theta_4) \big) + p_1 \big(1 - \sin(\phi_1 + \theta_1) \big) \Big) \, .
	\end{align}
There is a similar result for the anti-holomorphic sector. In general we see that in momentum space the vertices consist of delta function forms with insertions on external legs. Since the Map to AdS space, in momentum  space, is just a change of variables, this property persists. In any case there are no inverse Laplacians in the vertices, a problem that plaged other  constructions as discussed in
\cite{Sleight:2017pcz}.

\section{Bi-local Witten Diagrams}
\label{sec:bi-local witten diagrams}

The bi-local field $\Phi$ was observed to transform (through the momentum space map after Fourier transformation) into the physical AdS higher spin field. On the CFT side, the bi-local field also represents the generating function of higher spin currents for the free theory as
	\begin{equation}
		J_s(x, \epsilon) \, = \, \Big[ \dop \, \Phi(x_1, x_2) \Big]_{x_1=x_2=x} \, , 
	\end{equation}
with $\epsilon$ representing a polarization vector and (see for example~\cite{Skvortsov:2015pea,Giombi:2016hkj}) 
	\begin{equation}
		\dop \, \equiv \, \sum_{k=0}^{\infty} \frac{(-1)^k}{k! (s-k)!\Gamma(k-1+{d\over 2})\Gamma(s-k-1+{d\over 2})} \, (\epsilon \cdot \partial_{x_1})^{s-k} (\epsilon \cdot \partial_{x_2})^k \, .
	\label{D_s}
	\end{equation}
Consequently, all correlation functions of higher spin currents 
	\begin{equation}
		\Big\langle J_{s_1}(x_1, \epsilon_1) J_{s_2}(x_2, \epsilon_2) \cdots J_{s_n}(x_n, \epsilon_n) \Big\rangle
	\end{equation}
follow from correlators of bi-local fields
	\begin{equation}
		\Big\langle \Phi(x_1, x_1') \Phi(x_2, x_2') \cdots \Phi(x_n, x_n') \Big\rangle \, .
	\end{equation}
With this in mind, it is useful to explore the conformal block decomposition of the bi-local field theory. Indeed, the language of conformal blocks allows efficient and maximal use of the underlying conformal invariance. For recent progress in this direction see for example~\cite{Caron-Huot:2017vep,Alday:2017vkk,Giombi:2018vtc,Aharony:2018npf,Mukhametzhanov:2018zja}.

\subsection{Bi-local Propagator}
\label{sec:bi-local propagator}

We start with the evaluation of the propagator of  bi-local fluctuations
represented by $\etaflc$ 
\begin{equation}
    \cD(x_1,x_2;x_3,x_4)\equiv\langle \eta(x_1,x_2)\eta(x_3,x_4) \rangle \, .
\end{equation}
We follow the  ($\lambda=0$) case, noting the fact that the discussion of the  IR fixed point case is essentially the same (with only a boundary condition change for the lowest mode). A recent on-shell discussion of IR bi-local states is given in\cite{Mulokwe:2018czu}.

For the propagator we consider the linearized collective field equation, which is the Schwinger-Dyson equation for the two point function of the bi-local fluctuation $\etaflc$ at large $N$ 
\begin{align}
	\nabla_1\nabla_2 \cD(x_1,x_2;x_3,x_4)=\delta(x_{13}) \delta(x_{24})+\delta(x_{14}) \delta(x_{23})\, .\label{eq: sd eq of 2pt}
\end{align}
This equation is easily solved,with the propagator expressed as  
	\begin{equation}
		\cD(x_1, x_2;x_3, x_4) \, = \, \Phi_0(x_1, x_3) \Phi_0(x_2, x_4) \, + \, \Phi_0(x_1, x_4) \Phi_0(x_2, x_3) \, ,
	\end{equation}
which is diagrammatically represented by \eqref{diagram:propagator}. 
We denote the classical solution\footnote{This corresponds to the two point function of the free $O(N)$ scalar field. \ie ${1\over N}\sum_i \langle \phi_i(x_1)\phi_i(x_2) \rangle$.} $\Phi_0$ by a thick line as (the second equality is for $d=3$)
\begin{equation}
		\Phi_0(x_1,x_2)\;=\; 
		\begin{minipage}[c]{0.2\linewidth}
            \centering
            \begin{tikzpicture}
            \draw [line width=1.2pt] (-0.9,0)-- (0.9,0);
            \draw (-1.4,0.0) node {$x_1$};
            \draw (1.4,0.0) node {$x_2$};
            \draw [fill] (-0.9,0) circle [radius=.08];
            \draw [fill] (0.9,-0) circle [radius=.08];
            \end{tikzpicture} 
        \end{minipage}\hspace{5mm}=\;{1\over 4\pi}{1 \over |x_{12}| }\, .
\end{equation}
In addition, the inverse\footnote{$\Phi_0^{-1}(x_1,x_2)$ is the inverse of $\Phi_0(x_1,x_2)$ as a matrix in $(x_1,x_2)$ space.} $\Phi_0^{-1}$ of the classical solution, which corresponds to a differential operator, is represented by a dashed line
\begin{equation}
		\Phi_0^{-1}(x_1,x_2)\;=\;-\nabla_1^2 \delta^d(x_{12}) \;=\; -\nabla_2^2 \delta^d(x_{12}) \;=\;
		\begin{minipage}[c]{0.2\linewidth}
            \centering
            \begin{tikzpicture}
            \draw [line width=1.2pt, dashed] (-0.9,0)-- (0.9,0);
            \draw (-1.4,0.0) node {$x_1$};
            \draw (1.4,0.0) node {$x_2$};
            \draw [fill] (-0.9,0) circle [radius=.08];
            \draw [fill] (0.9,-0) circle [radius=.08];
            \end{tikzpicture} 
        \end{minipage} 
	\end{equation}
From the above definitions, the contraction of a single thick line and a single dash line is easily evaluated
\begin{equation}
		\Phi_0^{-1}(x_1,x_3)\star \Phi_0(x_3,x_2)\;=\;-\nabla_1^2\Phi_0(x_1,x_2) \;=\;
		\begin{minipage}[c]{0.2\linewidth}
            \centering
            \begin{tikzpicture}
            \draw [line width=1.2pt]  (0,0) -- (1.2,0);
            \draw [line width=1.2pt, dashed] (-1.2,0) -- (0,0) ;
            \draw (-1.7,0.0) node {$x_1$};
            \draw (0,-0.3) node {$x_3$};
            \draw (1.7,0.0) node {$x_2$};
            \draw [fill] (-1.2,0) circle [radius=.08];
            \draw [fill] (0,0) circle [radius=.08];
            \draw [fill] (1.2,-0) circle [radius=.08];
            \end{tikzpicture} 
        \end{minipage}\hspace{10mm} =\; \delta^d(x_{12})
	\end{equation}
The bi-local propagator of the $\etaflc$ field is given by 
	\begin{equation}
\cD(x_1,x_2;x_3,x_4)\equiv
\begin{minipage}[c]{0.2\linewidth}
\centering
    \begin{tikzpicture}
\draw [line width=1.2pt] (-0.9,0.2)-- (0.9,0.2);
\draw [line width=1.2pt] (-0.9,-0.2)-- (0.9,-0.2);
\draw [fill=black!40,opacity=0.4] (-0.9,0.2)-- (0.9,0.2) -- (0.9,-0.2)-- (-0.9,-0.2);
\draw (-1.28,0.27) node {$x_1$};
\draw (-1.28,-0.27) node {$x_2$};
\draw (1.28,0.27) node {$x_3$};
\draw (1.28,-0.27) node {$x_4$};
\draw [fill] (-0.9,0.2) circle [radius=.08];
\draw [fill] (-0.9,-0.2) circle [radius=.08];
\draw [fill] (0.9,0.2) circle [radius=.08];
\draw [fill] (0.9,-0.2) circle [radius=.08];
\end{tikzpicture} 
\end{minipage}
\equiv\;
\begin{minipage}[c]{0.2\linewidth}
\centering
    \begin{tikzpicture}
\draw [line width=1.2pt] (-0.9,0.2)-- (0.9,0.2);
\draw [line width=1.2pt] (-0.9,-0.2)-- (0.9,-0.2);
\draw (-1.28,0.27) node {$x_1$};
\draw (-1.28,-0.27) node {$x_2$};
\draw (1.28,0.27) node {$x_3$};
\draw (1.28,-0.27) node {$x_4$};
\draw [fill] (-0.9,0.2) circle [radius=.08];
\draw [fill] (-0.9,-0.2) circle [radius=.08];
\draw [fill] (0.9,0.2) circle [radius=.08];
\draw [fill] (0.9,-0.2) circle [radius=.08];
\end{tikzpicture} 
\end{minipage}
+  
\begin{minipage}[c]{0.2\linewidth}
\centering
    \begin{tikzpicture}
\draw [line width=1.2pt] plot [smooth] coordinates {(-0.9,0.2) (-0.3,0.2) (0,0) (0.3,-0.2) (0.9,-0.2)};
\draw [line width=1.2pt] plot [smooth] coordinates {(-0.9,-0.2) (-0.3,-0.2) (-0.1,-0.1)};
\draw [line width=1.2pt] plot [smooth] coordinates {(0.9,0.2) (0.3,0.2) (0.1,0.1) };
\draw (-1.28,0.27) node {$x_1$};
\draw (-1.28,-0.27) node {$x_2$};
\draw (1.28,0.27) node {$x_3$};
\draw (1.28,-0.27) node {$x_4$};
\draw [fill] (-0.9,0.2) circle [radius=.08];
\draw [fill] (-0.9,-0.2) circle [radius=.08];
\draw [fill] (0.9,0.2) circle [radius=.08];
\draw [fill] (0.9,-0.2) circle [radius=.08];
\end{tikzpicture} 
\end{minipage}
	\label{diagram:propagator}
	\end{equation}
This is the bilocal representation of the propagator.

Another representation of the propagator can be given in terms of conformal partial wave functions $\cpw_{\cdim,s}(x_1,\cdim_0;x_2,\cdim_0;y, -\cdim ,s,\varepsilon)$ in \eqref{eq: intro cpw} which we explain in detail in Appendix~\ref{app: conformal partial wave}
\begin{align}
	&\cD(x_1,x_2;x_3,x_4)\,\equiv\, \langle \eta(x_1,x_2)\eta(x_3,x_4) \rangle\cr
	=&\,\sum_{\cdim, s} \int d^dy \; {\rho_s(\cdim)\over 2}  \cpw_{\cdim,s}(x_1,\cdim_0;x_2,\cdim_0;y, -\cdim ,s,\partial_\epsilon) \cpw_{\cdim,s}(x_3,\cdim_0;x_4, \cdim_0 ;y,  \cdim ,s,\epsilon)
\end{align}
where the Plancherel weight $\rho_s(\cdim)$ is defined by
\begin{equation}
	\rho_s(\cdim)\, \equiv\, {\Gamma({d\over 2} +s)\over 2 (2\pi)^{d\over 2}s!}\left|{\Gamma(d/2-1+\cdim)\over \Gamma(\cdim)}\right|^2 [(d/2+s-1)^2-\cdim^2]\, .
\end{equation}
Note that $\sum_{\cdim,s}$ is composed of a summation over the even spins $s$ and a contour integral along the imaginary axis, with a deformation for $d\leqq 3$ (see \eqref{eq: def summation} for more details). 
%
%
One can easily see that the RHS of \eqref{eq: sd eq of 2pt} becomes the completeness relation of the conformal partial wave functions in~\eqref{eq: cpw completeness}, which proves \eqref{eq: sd eq of 2pt}.
%
%
%
The propagator has a physical pole at $c=s-{d-1\over 2}$. The
pole at $c=s+{d+1\over 2}$ is unphysical. It is clear that
\begin{align}
	& \nabla^2_1\nabla^2_2 \cD(x_1,x_2;x_3,x_4)=\hspace{5mm}\begin{minipage}[c]{0.2\linewidth}
\centering
    \begin{tikzpicture}
\draw [line width=1.2pt] (-0.9,0.2)-- (0.9,0.2);
\draw [line width=1.2pt] (-0.9,-0.2)-- (0.9,-0.2);
\draw [line width=1.2pt, dashed] (-0.9,0.2)-- (-1.9,0.2);
\draw [line width=1.2pt, dashed] (-0.9,-0.2)-- (-1.9,-0.2);
\draw [fill=black!40,opacity=0.4] (-0.9,0.2)-- (0.9,0.2) -- (0.9,-0.2)-- (-0.9,-0.2);
\draw (-0.9,0.6) node {1};
\draw (-0.9,-0.6) node {2};
\draw (1.2,0.24) node {3};
\draw (1.2,-0.24) node {4};
\draw [fill] (-0.9,0.2) circle [radius=.08];
\draw [fill] (-0.9,-0.2) circle [radius=.08];
\draw [fill] (0.9,0.2) circle [radius=.08];
\draw [fill] (0.9,-0.2) circle [radius=.08];
\end{tikzpicture} 
\end{minipage}\cr
	=&2\sum_{\cdim,s} \int d^dy {\lambda_{\cdim,s}  \rho_s(\cdim)\over  |x_{12}|^{4} }  \cpw_{\cdim,s}(x_1,\cdim_0;x_2,\cdim_0;y, -\cdim ,s,\partial_\epsilon) \cpw_{\cdim,s}(x_3,\cdim_0;x_4, \cdim_0 ;y,  \cdim ,s,\epsilon)\cr
	=& \delta(x_{13})\delta(x_{24})+ \delta(x_{14})\delta(x_{23})\, .\label{eq: two pt function identity1}
\end{align}
Finally, note that the propagator $\cD$ also satisfies the following equation, which is useful to simplify the diagrammatics
\begin{equation}
\begin{minipage}[c]{0.2\linewidth}
\centering
    \begin{tikzpicture}
\draw [line width=1.2pt] (-0.9,0.2)-- (0.9,0.2);
\draw [line width=1.2pt] (-0.9,-0.2)-- (0.9,-0.2);
\draw [line width=1.2pt, dashed] (-0.9,0.2)-- (-1.9,0.2);
\draw [fill=black!40,opacity=0.4] (-0.9,0.2)-- (0.9,0.2) -- (0.9,-0.2)-- (-0.9,-0.2);
\draw (-0.9,0.6) node {1};
\draw (-0.9,-0.6) node {2};
\draw (1.2,0.24) node {3};
\draw (1.2,-0.24) node {4};
\draw [fill] (-0.9,0.2) circle [radius=.08];
\draw [fill] (-0.9,-0.2) circle [radius=.08];
\draw [fill] (0.9,0.2) circle [radius=.08];
\draw [fill] (0.9,-0.2) circle [radius=.08];
\end{tikzpicture} 
\end{minipage}=-\nabla^2_1 \cD(x_1,x_2;x_3,x_4)=\delta(x_{13})\Phi_0(x_{24})+\delta(x_{14})\Phi_0(x_{23})\, .\label{eq: two pt function identity2}
\end{equation}
%

\subsection{Bi-local Three-point Function}
\label{sec:three-point function}

In this subsection, we consider the bi-local 3-point function
	\begin{align}
		\Big\langle \etaflc(x_1, x_1') \etaflc(x_2, x_2') \etaflc(x_3, x_3') \Big\rangle \, .
	\label{eta 3pt}
	\end{align}
Acting with the differential operator in \eqref{D_s} on each leg of this bi-local 3-point function and setting $x_i = x_i'$, we can obtain any higher spin current 3-point function.

From the large $N$ expansion of the collective action, the cubic vertex can be read off as follows
\begin{align}
	\mathcal{V}_3(x_1,x'_1;x_2,x'_2;x_3,x'_3)\,=\,{1\over  N^{1\over 2}}  \Phi_0^{-1}(x'_3,x_1)\Phi_0^{-1}(x'_1,x_2)\Phi_0^{-1}(x'_2,x_3)\, .
\end{align}
The bi-local 3-point function can be computed from the bi-local Witten diagram rules to find 
	\begin{align}
		& \Big\langle \etaflc(x_1, x_1') \etaflc(x_2, x_2') \etaflc(x_3, x_3') \Big\rangle \nonumber\\
		=& \, \int d^dy_a d^dy_a' d^dy_b d^dy_b' d^dy_c d^dy_c' \, \mathcal{V}_3(y_a, y_a'; y_b, y_b'; y_c, y_c') \nonumber\\
		& \qquad \times \cD(x_1, x_1'; y_a, y_a') \cD(x_2, x_2'; y_b, y_b') \cD(x_3, x_3'; y_c, y_c') \, .
	\end{align}
Therefore, the bi-local three-point function is now represented as
\begin{align}
&\Big\langle \eta(x_1,x'_1)\eta(x_2,x'_2)\eta(x_3,x'_3) \Big \rangle\cr
=&\hspace{5mm}\begin{minipage}[c]{0.2\linewidth}
\centering
    \begin{tikzpicture}
        \draw [line width=1.2pt, dashed,domain=-15:75] plot ({0.7*cos(\x)}, {0.7*sin(\x)});
        \draw [line width=1.2pt, dashed,domain=105:195] plot ({0.7*cos(\x)}, {0.7*sin(\x)});
        \draw [line width=1.2pt, dashed,domain=-45:-135] plot ({0.7*cos(\x)}, {0.7*sin(\x)});
        \draw [line width=1.2pt] ({0.7*cos(105)}, {0.7*sin(105)}) -- ({0.7*cos(105)}, {0.7*sin(105)+0.7});
        \draw [line width=1.2pt] ({0.7*cos(75)}, {0.7*sin(75)}) -- ({0.7*cos(75)}, {0.7*sin(75)+0.7});
        \draw [line width=1.2pt] ({0.7*cos(-15)}, {0.7*sin(-15)}) -- ({0.7*cos(-15)+0.7*cos(-30)}, {0.7*sin(-15)+0.7*sin(-30)});
        \draw [line width=1.2pt] ({0.7*cos(-45)}, {0.7*sin(-45)}) -- ({0.7*cos(-45)+0.7*cos(-30)}, {0.7*sin(-45)+0.7*sin(-30)});
        \draw [line width=1.2pt] ({0.7*cos(-135)}, {0.7*sin(-135)}) -- ({0.7*cos(-135)+0.7*cos(-150)}, {0.7*sin(-135)+0.7*sin(-150)});
        \draw [line width=1.2pt] ({0.7*cos(195)}, {0.7*sin(195)}) -- ({0.7*cos(195)+0.7*cos(-150)}, {0.7*sin(195)+0.7*sin(-150)});
        \draw [fill] ({0.7*cos(75)}, {0.7*sin(75)+0.7}) circle [radius=.08];
        \draw [fill] ({0.7*cos(105)}, {0.7*sin(105)+0.7}) circle [radius=.08];
        \draw [fill] ({0.7*cos(-15)+0.7*cos(-30)}, {0.7*sin(-15)+0.7*sin(-30)}) circle [radius=.08];
        \draw [fill] ({0.7*cos(-45)+0.7*cos(-30)}, {0.7*sin(-45)+0.7*sin(-30)}) circle [radius=.08];
        \draw [fill] ({0.7*cos(195) + 0.7*cos(-150)}, {0.7*sin(195)+ 0.7*sin(-150)}) circle [radius=.08];
        \draw [fill] ({0.7*cos(-135)+0.7*cos(-150)}, {0.7*sin(-135)+0.7*sin(-150)}) circle [radius=.08];
        \draw (0,0) node {\textit{3}};
        \draw ({0.7*cos(105)-0.3}, {0.7*sin(105)+0.7+0.2}) node {$x_1$};
        \draw ({0.7*cos(75)+0.3}, {0.7*sin(75)+0.7+0.2}) node {$x'_1$};
        \draw ({0.7*cos(-15)+0.7*cos(-30)+0.1}, {0.7*sin(-15)+0.7*sin(-30)+0.25}) node {$x_2$};
        \draw ({0.7*cos(-45)+0.7*cos(-30)+0.0},{0.7*sin(-45)+0.7*sin(-30)-0.3}) node {$x'_2$};
        \draw ({0.7*cos(-135)+0.7*cos(-150)}, {0.7*sin(-135)+0.7*sin(-150)-0.3}) node {$x_3$};
        \draw ({0.7*cos(195) + 0.7*cos(-150)-0.1}, {0.7*sin(195) + 0.7*sin(-150)+0.25}) node {$x'_3$};
        \draw [fill=black!40,opacity=0.4] ({0.7*cos(75)}, {0.7*sin(75)})-- ({0.7*cos(75)}, {0.7*sin(75)+0.7}) -- ({0.7*cos(105)}, {0.7*sin(105)+0.7}) -- ({0.7*cos(105)}, {0.7*sin(105)});
        \draw [fill=black!40,opacity=0.4] ({0.7*cos(-15)}, {0.7*sin(-15)}) -- ({0.7*cos(-15)+0.7*cos(-30)}, {0.7*sin(-15)+0.7*sin(-30)}) -- ({0.7*cos(-45)+0.7*cos(-30)}, {0.7*sin(-45)+0.7*sin(-30)}) -- ({0.7*cos(-45)}, {0.7*sin(-45)});
        \draw [fill=black!40,opacity=0.4] ({0.7*cos(-135)}, {0.7*sin(-135)}) -- ({0.7*cos(-135)+0.7*cos(-150)}, {0.7*sin(-135)+0.7*sin(-150)}) -- ({0.7*cos(195)+0.7*cos(-150)}, {0.7*sin(195)+0.7*sin(-150)}) -- ({0.7*cos(195)}, {0.7*sin(195)});
\end{tikzpicture} 
\end{minipage} \hspace{5mm}  =\hspace{5mm}  \begin{minipage}[c]{0.2\linewidth}
\centering
    \begin{tikzpicture}
        \draw [line width=1.2pt ] ({0.7*cos(75)}, {0.7*sin(75)+0.7})   to [bend angle = 30, bend right] ({0.7*cos(-15)+0.7*cos(-30)}, {0.7*sin(-15)+0.7*sin(-30)});
        \draw [line width=1.2pt ] ({0.7*cos(-45)+0.7*cos(-30)},{0.7*sin(-45)+0.7*sin(-30)})   to [bend angle = 30, bend right] ({0.7*cos(-135)+0.7*cos(-150)}, {0.7*sin(-135)+0.7*sin(-150)});
        \draw [line width=1.2pt ] ({0.7*cos(195) + 0.7*cos(-150)}, {0.7*sin(195) + 0.7*sin(-150)})   to [bend angle = 30, bend right] ({0.7*cos(105)}, {0.7*sin(105)+0.7});
        \draw [fill] ({0.7*cos(75)}, {0.7*sin(75)+0.7}) circle [radius=.08];
        \draw [fill] ({0.7*cos(105)}, {0.7*sin(105)+0.7}) circle [radius=.08];
        \draw [fill] ({0.7*cos(-15)+0.7*cos(-30)}, {0.7*sin(-15)+0.7*sin(-30)}) circle [radius=.08];
        \draw [fill] ({0.7*cos(-45)+0.7*cos(-30)}, {0.7*sin(-45)+0.7*sin(-30)}) circle [radius=.08];
        \draw [fill] ({0.7*cos(195) + 0.7*cos(-150)}, {0.7*sin(195)+ 0.7*sin(-150)}) circle [radius=.08];
        \draw [fill] ({0.7*cos(-135)+0.7*cos(-150)}, {0.7*sin(-135)+0.7*sin(-150)}) circle [radius=.08];
        \draw ({0.7*cos(105)-0.3}, {0.7*sin(105)+0.7+0.2}) node {$x_1$};
        \draw ({0.7*cos(75)+0.3}, {0.7*sin(75)+0.7+0.2}) node {$x_1'$};
        \draw ({0.7*cos(-15)+0.7*cos(-30)+0.1}, {0.7*sin(-15)+0.7*sin(-30)+0.25}) node {$x_2$};
        \draw ({0.7*cos(-45)+0.7*cos(-30)+0.0},{0.7*sin(-45)+0.7*sin(-30)-0.3}) node {$x_2'$};
        \draw ({0.7*cos(-135)+0.7*cos(-150)}, {0.7*sin(-135)+0.7*sin(-150)-0.3}) node {$x_3$};
        \draw ({0.7*cos(195) + 0.7*cos(-150)-0.1}, {0.7*sin(195) + 0.7*sin(-150)+0.25}) node {$x_3'$};
\end{tikzpicture} 
\end{minipage}\quad+\quad \mbox{(7 permutations)}
\end{align}
where we used \eqref{eq: two pt function identity2}, which gives the permutation terms in each $x_i \leftrightarrow x_i'$ with a symmetry factor ${1\over 2}$. Hence, the three point function of the bi-local fluctuation is given by 
	\begin{align}
		& \Big\langle \etaflc(x_1, x_1') \etaflc(x_2, x_2') \etaflc(x_3, x_3') \Big\rangle \nonumber\\
		=& \, {1\over \sqrt{N} } {{1\over (4\pi)^3}\over |x_1' - x_2|^{2\Delta_0}|x_2' - x_3|^{2\Delta_0}|x_3' - x_1|^{2\Delta_0}} \, + \, ({\rm 7\ permutations}) \, .
	\end{align}

\subsection{Bi-local $s$-channel Contribution}
\label{sec:s-channel}

\begin{figure}[t!]
\centering
\begin{minipage}[c]{0.05\linewidth}
\centering
$\widehat{A}\;\;= $
\end{minipage}
\begin{minipage}[c]{0.23\linewidth}
\centering
        \begin{tikzpicture}

\def\tikzx{1.2}
\def\tikzd{0.3}

\draw [fill] (\tikzx,\tikzx) circle [radius=.08];
\draw [fill] (-\tikzx,\tikzx) circle [radius=.08];
\draw [fill] (\tikzx,-\tikzx) circle [radius=.08];
\draw [fill] (-\tikzx,-\tikzx) circle [radius=.08];

\draw [fill] (\tikzx,\tikzx-\tikzd) circle [radius=.08];
\draw [fill] (-\tikzx,\tikzx-\tikzd) circle [radius=.08];
\draw [fill] (\tikzx,-\tikzx+\tikzd) circle [radius=.08];
\draw [fill] (-\tikzx,-\tikzx+\tikzd) circle [radius=.08];

\draw [line width=1.2pt] (\tikzx,\tikzx) -- (-\tikzx,\tikzx);
\draw [line width=1.2pt] (\tikzx,-\tikzx) -- (-\tikzx,-\tikzx);

\draw [line width=1.2pt] (\tikzx,-\tikzx+\tikzd) -- (\tikzx,\tikzx-\tikzd);
\draw [line width=1.2pt] (-\tikzx,-\tikzx+\tikzd) -- (-\tikzx,\tikzx-\tikzd);
\end{tikzpicture}
\end{minipage}
\quad
\begin{minipage}[c]{0.05\linewidth}
\centering
$\widehat{B}\;\;= $
\end{minipage}
\begin{minipage}[c]{0.23\linewidth}
\centering
        \begin{tikzpicture}

\def\tikzx{1.2}
\def\tikzd{0.3}

\draw [fill] (\tikzx,\tikzx) circle [radius=.08];
\draw [fill] (-\tikzx,\tikzx) circle [radius=.08];
\draw [fill] (\tikzx,-\tikzx) circle [radius=.08];
\draw [fill] (-\tikzx,-\tikzx) circle [radius=.08];

\draw [fill] (\tikzx-\tikzd,\tikzx) circle [radius=.08];
\draw [fill] (-\tikzx+\tikzd,\tikzx) circle [radius=.08];
\draw [fill] (\tikzx-\tikzd,-\tikzx) circle [radius=.08];
\draw [fill] (-\tikzx+\tikzd,-\tikzx) circle [radius=.08];

\draw [line width=1.2pt] (\tikzx,\tikzx) -- (\tikzx,-\tikzx);
\draw [line width=1.2pt] (-\tikzx,\tikzx) -- (-\tikzx,-\tikzx);

\draw [line width=1.2pt] (\tikzx-\tikzd,-\tikzx) -- (-\tikzx+\tikzd,\tikzx);
\draw [line width=1.2pt] (-\tikzx+\tikzd,-\tikzx) -- (\tikzx-\tikzd,\tikzx);
\end{tikzpicture}
\end{minipage}
\quad
\begin{minipage}[c]{0.05\linewidth}
\centering
$\widehat{C}\;\;= $
\end{minipage}
\begin{minipage}[c]{0.23\linewidth}
\centering
        \begin{tikzpicture}

\def\tikzx{1.2}
\def\tikzd{0.3}

\draw [fill] (\tikzx,\tikzx) circle [radius=.08];
\draw [fill] (-\tikzx,\tikzx) circle [radius=.08];
\draw [fill] (\tikzx,-\tikzx) circle [radius=.08];
\draw [fill] (-\tikzx,-\tikzx) circle [radius=.08];

\draw [fill] (\tikzx,\tikzx-\tikzd) circle [radius=.08];
\draw [fill] (-\tikzx,\tikzx-\tikzd) circle [radius=.08];
\draw [fill] (\tikzx,-\tikzx+\tikzd) circle [radius=.08];
\draw [fill] (-\tikzx,-\tikzx+\tikzd) circle [radius=.08];

\draw [line width=1.2pt] (\tikzx,\tikzx) -- (-\tikzx,\tikzx);
\draw [line width=1.2pt] (\tikzx,-\tikzx) -- (-\tikzx,-\tikzx);

\draw [line width=1.2pt] (\tikzx,-\tikzx+\tikzd) -- (-\tikzx,\tikzx-\tikzd);
\draw [line width=1.2pt] (-\tikzx,-\tikzx+\tikzd) --  (\tikzx,\tikzx-\tikzd);
\end{tikzpicture}
\end{minipage}

	\caption{Reduced Diagrams. Each channel is reduced as $s=2(\hat A+\hat B)$, $t=2(\hat A+\hat C)$ and $u=2(\hat B+\hat C$).}
	\label{fig:ABC}
\end{figure}
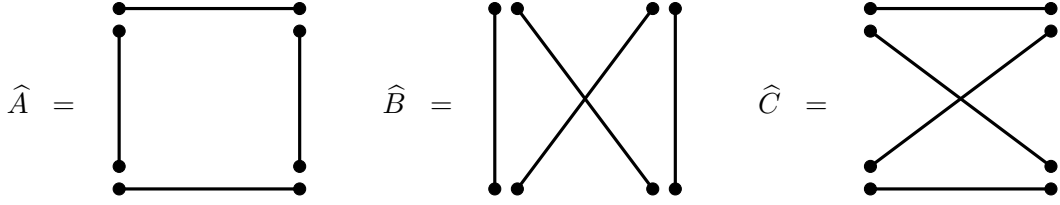

\begin{figure}[t]
\centering
	\begin{tikzpicture}[scale=1.2]
\draw [line width=1.2pt] (0.483,0.2)-- (1.29,1.007);
\draw [line width=1.2pt] (0.2,0.483)-- (1.007,1.29);
\draw [line width=1.2pt] (-0.483,0.2)-- (-1.29,1.007);
\draw [line width=1.2pt] (-0.2,0.483)-- (-1.007,1.29);
\draw [line width=1.2pt] (0.483,-0.2)-- (1.29,-1.007);
\draw [line width=1.2pt] (0.2,-0.483)-- (1.007,-1.29);
\draw [line width=1.2pt] (-0.483,-0.2)-- (-1.29,-1.007);
\draw [line width=1.2pt] (-0.2,-0.483)-- (-1.007,-1.29);
%
%
\draw [line width=1.2pt, dashed] (0.2,0.483)  to [bend angle = 30, bend right] (-0.2,0.483);
\draw [line width=1.2pt, dashed] (-0.2,-0.483)  to [bend angle = 30, bend right] (0.2,-0.483);
\draw [line width=1.2pt, dashed] (0.483,0.2)  to [bend angle = 30, bend left] (0.483,-0.2);
\draw [line width=1.2pt, dashed] (-0.483,0.2)  to [bend angle = 30, bend right] (-0.483,-0.2);
\draw [fill=black!40,opacity=0.4] (0.483,0.2)-- (1.29,1.007) -- (1.007,1.29) -- (0.2,0.483);
\draw [fill=black!40,opacity=0.4] (-0.483,0.2)-- (-1.29,1.007) -- (-1.007,1.29) -- (-0.2,0.483);
\draw [fill=black!40,opacity=0.4] (0.483,-0.2)-- (1.29,-1.007) -- (1.007,-1.29) -- (0.2,-0.483);
\draw [fill=black!40,opacity=0.4] (-0.483,-0.2)-- (-1.29,-1.007) -- (-1.007,-1.29) -- (-0.2,-0.483);
\draw [fill] (1.29,1.007) circle [radius=.08];
\draw [fill] (1.007,1.29) circle [radius=.08];
\draw [fill] (-1.29,1.007) circle [radius=.08];
\draw [fill] (-1.007,1.29) circle [radius=.08];
\draw [fill] (1.29,-1.007) circle [radius=.08];
\draw [fill] (1.007,-1.29) circle [radius=.08];
\draw [fill] (-1.29,-1.007) circle [radius=.08];
\draw [fill] (-1.007,-1.29) circle [radius=.08];
\draw (0,0) node {\textit{4}};
\end{tikzpicture}
	\caption{4-point contact diagram, which is reduced to $-3\hat A-3\hat B-3\hat C$ due to the cyclic symmetry of the vertex.}
	\label{fig:4}
\end{figure}
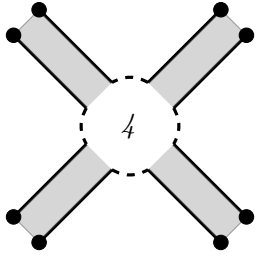

Now we evaluate the $s$-channel contribution, which is given by
\begin{align}
	&\mathcal{A}_s(11'22'33'44')\equiv \langle \eta(x_1,x'_1)\eta(x_2,x'_2)\eta(x_3,x'_3)\eta(x_4,x'_4) \rangle_s \cr
	=&\int \left[\prod_{i=1}^6dy_i dy'_i \right] \mathcal{V}_3(y_1,y'_1;y_2,y'_2;y_3,y'_3)\mathcal{V}_3(y_4,y'_4;y_5,y'_5;y_6,y'_6)\cr
	&\!\!\!\times \cD(y_3,y_3';y_4,y_4')\cD(x_1,x_1';y_1,y_1')\cD(x_2,x_2';y_2,y_2')\cD(x_3,x_3';y_5,y_5')\cD(x_4,x_4';y_6,y_6')
\end{align}
and similarly for $\mathcal{A}_t(11'22'33'44')$ and $\mathcal{A}_u(11'22'33'44')$.
From the diagrammatic rules (\ie \eqref{eq: two pt function identity1} and \eqref{eq: two pt function identity2}), the $s$-channel contribution is given by 
\begin{equation}
s\;= 
\begin{minipage}[c]{0.4\linewidth}
\centering
        \begin{tikzpicture}[scale=1.2]
\draw [line width=1.2pt] (-1.65,1.26)-- (-1.15,0.4);
\draw [line width=1.2pt] (-1.49,0.2)-- (-1.99,1.06);
\draw [line width=1.2pt] (-1.65,-1.26)-- (-1.15,-0.4);
\draw [line width=1.2pt] (-1.49,-0.2)-- (-1.99,-1.06);
\draw [line width=1.2pt] (1.65,1.26)-- (1.15,0.4);
\draw [line width=1.2pt] (1.49,0.2)-- (1.99,1.06);
\draw [line width=1.2pt] (1.65,-1.26)-- (1.15,-0.4);
\draw [line width=1.2pt] (1.49,-0.2)-- (1.99,-1.06);
\draw [line width=1.2pt] (-0.75,0.2)-- (0.75,0.2);
\draw [line width=1.2pt] (-0.75,-0.2)-- (0.75,-0.2);
\draw [line width=1.2pt, dashed] (-1.15,0.4)  to [bend angle = 30, bend left] (-0.75,0.2);
\draw [line width=1.2pt, dashed] (-1.49,0.2)  to [bend angle = 30, bend right] (-1.49,-0.2);
\draw [line width=1.2pt, dashed] (-1.15,-0.4)  to [bend angle = 30, bend right] (-0.75,-0.2);
\draw [line width=1.2pt, dashed] (1.15,0.4)  to [bend angle = 30, bend right] (0.75,0.2);
\draw [line width=1.2pt, dashed] (1.49,0.2)  to [bend angle = 30, bend left] (1.49,-0.2);
\draw [line width=1.2pt, dashed] (1.15,-0.4)  to [bend angle = 30, bend left] (0.75,-0.2);
\draw [fill] (-1.65,1.26) circle [radius=.08];
\draw [fill] (-1.99,1.06) circle [radius=.08];
\draw [fill] (1.65,1.26) circle [radius=.08];
\draw [fill] (1.99,1.06) circle [radius=.08];
\draw [fill] (1.65,-1.26) circle [radius=.08];
\draw [fill] (1.99,-1.06) circle [radius=.08];
\draw [fill] (-1.65,-1.26) circle [radius=.08];
\draw [fill] (-1.99,-1.06) circle [radius=.08];
\draw [fill=black!40,opacity=0.4] (-1.65,1.26)-- (-1.15,0.4) -- (-1.49,0.2)-- (-1.99,1.06);
\draw [fill=black!40,opacity=0.4] (1.65,1.26)-- (1.15,0.4) -- (1.49,0.2)-- (1.99,1.06);
\draw [fill=black!40,opacity=0.4] (-1.65,-1.26)-- (-1.15,-0.4) -- (-1.49,-0.2)-- (-1.99,-1.06);
\draw [fill=black!40,opacity=0.4] (1.65,-1.26)-- (1.15,-0.4) -- (1.49,-0.2)-- (1.99,-1.06);
\draw [fill=black!40,opacity=0.4] (-0.75,0.2)-- (0.75,0.2) -- (0.75,-0.2)-- (-0.75,-0.2);
\draw (-1.65+0.3,1.26+0.2) node {$x_1$};
\draw (-1.99-0.35,1.06-0.0) node {$x'_1$};
\draw (-1.99-0.4,-1.06) node {$x_2$};
\draw (-1.65+0.3,-1.26-0.2) node {$x'_2$};
\draw (1.99+0.35,1.06) node {$x_3$};
\draw (1.65-0.3,1.26+0.2) node {$x'_3$};
\draw (1.65-0.3,-1.26-0.2) node {$x_4$};
\draw (1.99+0.35,-1.06) node {$x'_4$};
\draw (-1.13,0) node {\textit{3}};
\draw (1.13,0) node {\textit{3}};
\end{tikzpicture}
\end{minipage}
\end{equation}
and it can be written as
\begin{equation}
	\mathcal{A}_s(1234)\,\equiv\, \langle \eta(x_1,x'_1)\eta(x_2,x'_2)\eta(x_3,x'_3)\eta(x_4,x'_4) \rangle_s=(\hat A+\hat B)\label{eq: s channel general}
\end{equation}
where $\hat A$, $\hat B$ and $\hat C$ are defined in Figure \ref{fig:ABC}.
Similarly the $t$-channel contribution is given by 
\begin{equation}
    t\;= 
\begin{minipage}[c]{0.4\linewidth}
\centering
        \begin{tikzpicture}
\draw [line width=1.2pt] (1.26,-1.65)-- (0.4,-1.15);
\draw [line width=1.2pt] (0.2,-1.49)-- (1.06,-1.99);
\draw [line width=1.2pt] (-1.26,-1.65)-- (-0.4,-1.15);
\draw [line width=1.2pt] (-0.2,-1.49)-- (-1.06,-1.99);
\draw [line width=1.2pt] (1.26,1.65)-- (0.4,1.15);
\draw [line width=1.2pt] (0.2,1.49)-- (1.06,1.99);
\draw [line width=1.2pt] (-1.26,1.65)-- (-0.4,1.15);
\draw [line width=1.2pt] (-0.2,1.49)-- (-1.06,1.99);
\draw [line width=1.2pt] (0.2,-0.75)-- (0.2,0.75);
\draw [line width=1.2pt] (-0.2,-0.75)-- (-0.2,0.75);
\draw [line width=1.2pt, dashed] (0.4,-1.15)  to [bend angle = 30, bend right] (0.2,-0.75);
\draw [line width=1.2pt, dashed] (0.2,-1.49)  to [bend angle = 30, bend left] (-0.2,-1.49);
\draw [line width=1.2pt, dashed] (-0.4,-1.15)  to [bend angle = 30, bend left] (-0.2,-0.75);
\draw [line width=1.2pt, dashed] (0.4,1.15)  to [bend angle = 30, bend left] (0.2,0.75);
\draw [line width=1.2pt, dashed] (0.2,1.49)  to [bend angle = 30, bend right] (-0.2,1.49);
\draw [line width=1.2pt, dashed] (-0.4,1.15)  to [bend angle = 30, bend right] (-0.2,0.75);
\draw [fill] (1.26,-1.65) circle [radius=.08];
\draw [fill] (1.06,-1.99) circle [radius=.08];
\draw [fill] (1.26,1.65) circle [radius=.08];
\draw [fill] (1.06,1.99) circle [radius=.08];
\draw [fill] (-1.26,1.65) circle [radius=.08];
\draw [fill] (-1.06,1.99) circle [radius=.08];
\draw [fill] (-1.26,-1.65) circle [radius=.08];
\draw [fill] (-1.06,-1.99) circle [radius=.08];
\draw [fill=black!40,opacity=0.4] (1.26,-1.65)-- (0.4,-1.15) -- (0.2,-1.49)-- (1.06,-1.99);
\draw [fill=black!40,opacity=0.4] (1.26,1.65)-- (0.4,1.15) -- (0.2,1.49)-- (1.06,1.99);
\draw [fill=black!40,opacity=0.4] (-1.26,-1.65)-- (-0.4,-1.15) -- (-0.2,-1.49)-- (-1.06,-1.99);
\draw [fill=black!40,opacity=0.4] (-1.26,1.65)-- (-0.4,1.15) -- (-0.2,1.49)-- (-1.06,1.99);
\draw [fill=black!40,opacity=0.4] (0.2,-0.75)-- (0.2,0.75) -- (-0.2,0.75)-- (-0.2,-0.75);
\draw (-1.06,1.99+0.35) node {$x_1$};
\draw (-1.26-0.2,1.65-0.3) node {$x'_1$};
\draw (-1.26-0.2,-1.65+0.3) node {$x_2$};
\draw (-1.06,-1.99-0.4) node {$x'_2$};
\draw (1.26+0.2,1.65-0.3) node {$x_3$};
\draw (1.06,1.99+0.35) node {$x'_3$};
\draw (1.06-0.0,-1.99-0.35) node {$x_4$};
\draw (1.26+0.2,-1.65+0.3) node {$x'_4$};
\draw (0,-1.13) node {\textit{3}};
\draw (0,1.13) node {\textit{3}};
\end{tikzpicture}
\end{minipage}
\end{equation} 
\begin{equation}
	\mathcal{A}_t(1234)\,\equiv\, \langle \eta(x_1,x'_1)\eta(x_2,x'_2)\eta(x_3,x'_3)\eta(x_4,x'_4) \rangle_t =2(\hat A+\hat C)\label{eq: t channel general}
\end{equation}
and the $u$-channel contribution is

\begin{equation}
u\;= 
\begin{minipage}[c]{0.4\linewidth}
\centering
        \begin{tikzpicture}[scale=1.2]
\draw[double=black!20,fill opacity=0.4,double distance=0.4cm,line join=round,looseness=0.5,line width=1.2pt] (1.99,1.06) to[out=180,in=120] (-1.32,0.3);
\draw[double=black!20,fill opacity=0.4,double distance=0.4cm,line join=round,looseness=0.5,line width=1.2pt] (-1.99,1.06) to[out=0,in=60] (1.32,0.3);
\draw [line width=1.2pt] (-1.65,-1.26)-- (-1.15,-0.4);
\draw [line width=1.2pt] (-1.49,-0.2)-- (-1.99,-1.06);
\draw [line width=1.2pt] (1.65,-1.26)-- (1.15,-0.4);
\draw [line width=1.2pt] (1.49,-0.2)-- (1.99,-1.06);
\draw [line width=1.2pt] (-0.75,0.2)-- (0.75,0.2);
\draw [line width=1.2pt] (-0.75,-0.2)-- (0.75,-0.2);
\draw [line width=1.2pt, dashed] (-1.15,0.4)  to [bend angle = 30, bend left] (-0.75,0.2);
\draw [line width=1.2pt, dashed] (-1.49,0.2)  to [bend angle = 30, bend right] (-1.49,-0.2);
\draw [line width=1.2pt, dashed] (-1.15,-0.4)  to [bend angle = 30, bend right] (-0.75,-0.2);
\draw [line width=1.2pt, dashed] (1.15,0.4)  to [bend angle = 30, bend right] (0.75,0.2);
\draw [line width=1.2pt, dashed] (1.49,0.2)  to [bend angle = 30, bend left] (1.49,-0.2);
\draw [line width=1.2pt, dashed] (1.15,-0.4)  to [bend angle = 30, bend left] (0.75,-0.2);
\draw [fill] (-1.99,1.26) circle [radius=.08];
\draw [fill] (-1.99,0.86) circle [radius=.08];
\draw [fill] (1.98,1.26) circle [radius=.08];
\draw [fill] (1.99,0.86) circle [radius=.08];
\draw [fill] (1.65,-1.26) circle [radius=.08];
\draw [fill] (1.99,-1.06) circle [radius=.08];
\draw [fill] (-1.65,-1.26) circle [radius=.08];
\draw [fill] (-1.99,-1.06) circle [radius=.08];
\draw [fill=black!40,opacity=0.4] (-1.65,-1.26)-- (-1.15,-0.4) -- (-1.49,-0.2)-- (-1.99,-1.06);
\draw [fill=black!40,opacity=0.4] (1.65,-1.26)-- (1.15,-0.4) -- (1.49,-0.2)-- (1.99,-1.06);
\draw [fill=black!40,opacity=0.4] (-0.75,0.2)-- (0.75,0.2) -- (0.75,-0.2)-- (-0.75,-0.2);
\draw (-1.99,1.26+0.25) node {$x_1$};
\draw (-1.99,0.86-0.3) node {$x'_1$};
\draw (-1.99-0.4,-1.06) node {$x_2$};
\draw (-1.65+0.3,-1.26-0.2) node {$x'_2$};
\draw (1.99,0.86-0.3) node {$x_3$};
\draw (1.99,1.26+0.25) node {$x'_3$};
\draw (1.65-0.3,-1.26-0.2) node {$x_4$};
\draw (1.99+0.35,-1.06) node {$x'_4$};
\draw (-1.13,0) node {\textit{3}};
\draw (1.13,0) node {\textit{3}};
\end{tikzpicture}
\end{minipage}
\end{equation} 
\begin{equation}
	\mathcal{A}_u(1234)\,\equiv\, \langle \eta(x_1,x'_1)\eta(x_2,x'_2)\eta(x_3,x'_3)\eta(x_4,x'_4) \rangle_u=2(\hat B+\hat C)\, .\label{eq: u channel general}
\end{equation}
The quartic contribution associated with the four vertex
\begin{align}
	\mathcal{V}_4(x_1,x'_1;x_2,x'_2;x_3,x'_3;x_4,x_4')= - {1\over  N}  \Phi_0^{-1}(x'_3,x_1)\Phi_0^{-1}(x'_1,x_2)\Phi_0^{-1}(x'_2,x_3)\Phi_0^{-1}(x'_2,x_3)\label{eq: quartic vertex}
\end{align}
from Figure~\ref{fig:4} 
gives
\begin{equation}
	\mathcal{A}_4(1234)\equiv \langle \eta(x_1,x'_1)\eta(x_2,x'_2)\eta(x_3,x'_3)\eta(x_4,x'_4) \rangle_4=-3(\hat A+\hat B+\hat C)\, .\label{eq: quartic general}
\end{equation}
We will now evaluate the same $s$, $t$, $u$ channel diagrams using the conformal partial wave basis.
We will consider the four point amplitude of four scalars, even though the evaluation of the general case is almost identical in the bilocal approach \ie we consider the four point amplitude of
\begin{equation}
    \mathcal{O}(x)\,\equiv\, \lim_{x'\rightarrow x} \eta(x,x')\,=\,{1 \over \sqrt{N} }\sum_{i=1}^N : \phi_i(x)\phi_i(x) :\, .
\end{equation}
Note that with this definition the two point function of the scalar operator $\mathcal{O}(x)$ is given by
\begin{equation}
    \langle \mathcal{O}(x_1)\mathcal{O}(x_2)  \rangle\,=\, {1\over 8 \pi^2} {1 \over |x_{12}|^{4\Delta_0}}\,\equiv\, {\normalization \over |x_{12}|^{4\Delta_0}}\, .\label{eq: two point ftn eta}
\end{equation}
By setting $y'_i= y_i$ ($i=1,2,3,4$) in \eqref{eq: s channel general}, one obtains the $s$-channel correlation function of the four scalar $\mathcal{O}$ operators
\begin{align}
	&\mathcal{A}_s(1234)\equiv  \left.\langle \eta(x_1,x'_1)\eta(x_2,x'_2)\eta(x_3,x'_3)\eta(x_4,x'_4) \rangle_s\right|_{x'_i\rightarrow x_i} \cr
	=&{16\over N} {{1\over 2}\Upsilon_{\mathcal{O}}\over |x_{12}|^{2\Delta_0}|x_{34}|^{2\Delta_0}} \cD(x_1,x_2;x_3,x_4)\, .\label{eq: s channel1}
\end{align}
At coincident points $\hat{A},\hat{B},\hat{C}$ becomes $4A,4B,4C$. To evaluate the $s$-channel contribution, use the integral representation of the conformal block given in \cite{Dolan:2011dv,Bekaert:2015tva} (see also \cite{SimmonsDuffin:2012uy})
\begin{align}
	&\int d^dx {|x_{12}|^s|x_{34}|^s\widehat{C}_s\left( {x_{12}\cdot x_{34}\over |x_{12}||x_{34}| } \right) \over  |x-x_1|^{{d\over 2}+\cdim-s} |x-x_2|^{{d\over 2}+\cdim-s} |x-x_3|^{{d\over 2}-\cdim-s} |x-x_4|^{{d\over 2}-\cdim-s} }\cr
	=&\,{1\over |x_{12}|^{{d\over 2}+\cdim-s} |x_{34}|^{{d\over 2}-\cdim-s} } \left[K_{\cdim,s}G_{{d\over 2}+\cdim,s}(u,v)+K_{-\cdim,s}G_{{d\over 2}-\cdim,s}(u,v) \right]\label{eq: integral representation of block}
\end{align}
where the conformal cross ratios are given by
\begin{equation}
		u \, \equiv \, \frac{x_{12}^2 x_{34}^2}{x_{13}^2 x_{24}^2} \quad , \qquad v \, \equiv \, \frac{x_{14}^2 x_{23}^2}{x_{13}^2 x_{24}^2} \, .
\end{equation}
$\widehat{C}_s(x)$ is a Gegenbauer polynomial\footnote{The relation to the usual Gegenbauer polynomial $C_s(x)$ is given by
\begin{equation}
    \widehat{C}_s\left({X\cdot Y\over |X| |Y|}\right)= {s!\over 2^s({d\over 2}-1)_s}C_s\left({X\cdot Y\over |X| |Y|}\right)={(X^{\mu_1} \cdots X^{\mu_s}-\text{trace})(Y_{\mu_1} \cdots Y_{\mu_s}-\text{trace})\over |X|^s |Y|^s}\, .
\end{equation}} defined in~\cite{Simmons-Duffin:2017nub} and the coefficient $K_{c,s}$ is given by
\begin{equation}
	K_{\cdim,s}\,=\,{\pi^{1+{d\over 2}} 2^{4-d-2\cdim-2s} \Gamma(-\cdim) \Gamma({d-2\over 2}-\cdim+s)\Gamma({d-2\over 2}+\cdim+s)\over (\cdim+s+{d-2\over 2}) \Gamma({d-2\over 2}-\cdim) \left[\Gamma({{d\over 2} -\cdim+s \over 2})\Gamma({{d-2\over 2} + \cdim +s \over 2})\right]^2}\, .
\end{equation}
We follow the conventions of\footnote{Here is the comparison between different conventions
\begin{equation}
	G_{h,s}^{\text{ \cite{Penedones:2015aga}}}(u,v)={4^h  (d-2)_s\over (-2)^s ({d-2\over 2})_s}G^{\text{\cite{Kos:2014bka}}}_{h,s}(u,v)= {\Gamma(s+d-2))\Gamma({d-2\over 2}) \over 2^s \Gamma(d-s) \Gamma(s+{d-2\over 2})}G^{\text{\cite{Dolan:2011dv}}}_{h,s}(u,v)\, . \label{eq: convention of blocks}
\end{equation}} \cite{Penedones:2015aga} for the conformal block $G_{h,s}(u,v)$. Using this integral representation, one can write the $s$-channel contribution in \eqref{eq: s channel1} in terms of the conformal blocks. Therefore, we have (specializing for $d=3$)
\begin{align}
	\mathcal{A}_s(1234)\,=&\,{16\over N} {{1\over 2}\Upsilon_{\mathcal{O}}\over |x_{12}||x_{34}|}  \cD(x_1,x_2;x_3,x_4)\cr
	=&\, {1\over  |x_{12}|^{2}|x_{34}|^{2}}{32\over N} \Upsilon_{\mathcal{O}} \sum_{\cdim,s} {2^{{5\over 2}+s}\over (2\pi)^3}{\rho_s(\cdim)  \over  \lambda_{\cdim,s}}K_{\cdim,s}G_{{3\over 2}+\cdim,s}(u,v)\ ,\label{eq: s channel 2}
\end{align}
and
\begin{equation}
	 \lambda_{\cdim,s}\,=\, {1\over 4}
	 \left[ \cdim^2 - (s-{1\over 2})^2 \right]\left[ \cdim^2 - (s + {3\over 2})^2 \right]\, .
\end{equation}
%

%
Also, because $\rho_s(\cdim)$ and $\lambda_{h,s}(\cdim)$ in \eqref{eq: s channel1} are even functions of $\cdim$, the contribution of the conformal block and its shadow in \eqref{eq: integral representation of block} to the $s$-channel is identical.

By closing the contour in \eqref{eq: s channel 2} (see the contour defined in~\eqref{eq: def summation}) in the region $\text{Re}(\cdim)>0$, the $s$-channel $\mathcal{A}_s(1234)$ contribution picks up the following three types of simple poles.
\begin{itemize}
    \item Simple poles at $\cdim = s-{1\over 2}$ ( $s=0,2,4,\cdots$)\footnote{For $s=0$, this comes from the deformation of the contour, or equivalently from additional residue contributions. See~\eqref{eq: def summation}. } from the factor ${\rho_s(\cdim)  \over  \lambda_{\cdim,s}}K_{\cdim,s}$
    
    \item Simple poles at $\cdim = s+2n +{5\over 2}$ ($n=0,1,2,\cdots$ and $s=0,2,4,\cdots$) from the factor ${\rho_s(\cdim)  \over  \lambda_{\cdim,s}}K_{\cdim,s}$
    
    \item Simple poles at $\cdim = s-k+{1\over 2}$ ($k=1,2,\cdots, s$ and $s=0,2,4,\cdots$) from the conformal block $G_{h,s}(u,v)$
    
\end{itemize}
The first two types of simple poles can easily be seen in the expression for ${\rho_s(\cdim)  \over  \lambda_{\cdim,s}}K_{\cdim,s}$. On the other hand, the third type of simple poles in the conformal block $G_{h,s}(u,v)$ comes from null states. 

In Appendix~\ref{app: identity}, we show that the contributions from the second and the third types of poles cancel exactly. The remaining contribution to the $s$-channel comes from the first type of simple poles at $\cdim =s- {1\over 2}$ corresponding to single-trace operators. This leads to
\begin{align}
	\mathcal{A}_s(1234)=&-{1\over  |x_{12}|^{2}|x_{34}|^{2}} {32\over N} \Upsilon_{\mathcal{O}} \sum_{s} \res_{\cdim=s-{1\over 2}} {2^{{5\over 2}+s}\over (2\pi)^3}{\rho_s(\cdim)  \over  \lambda_{\cdim,s}}K_{\cdim,s}G_{{3\over 2}+\cdim,s}(u,v)\cr
	=&{2 \normalization^2 \over |x_{12}|^2|x_{34}|^2} \sum_{s} c_s^2 G_{s+1,s}(u,v)\label{eq: s channel 3p}
\end{align}
agreeing with the $s$-channel OPE expressions (for $2A+2B$). Here we define $A, B$ and $C$ as follows
\begin{align}
    A\, \equiv &\,\left.\hat{A}\right|_{x'_i\rightarrow x_i}\, =\, \frac{{4\over N}\Upsilon_{\mathcal{O}}^2}{(x_{12}^2 x_{34}^2)^{2\Delta_0}}  u^{\Delta_0}\, , \\
    B\,\equiv&\,\left.\hat{B}\right|_{x'_i\rightarrow x_i}\,=\,\frac{{4\over N}\Upsilon_{\mathcal{O}}^2}{(x_{12}^2 x_{34}^2)^{2\Delta_0}}  \left( \frac{u}{v} \right)^{\Delta_0}\, ,\\
    C\,\equiv&\,\left.\hat{C}\right|_{x'_i\rightarrow x_i}\,=\,\frac{{4\over N}\Upsilon_{\mathcal{O}}^2}{(x_{12}^2 x_{34}^2)^{2\Delta_0}} \left( \frac{u}{v} \right)^{\Delta_0}u^{\Delta_0}\, .
\end{align}
Note that we include a normalization factor $\Upsilon_{\mathcal{O}}^2$. A summary is given in the Appendix \ref{app:spectrum form of the bi-local propagator}.

The $t$-channel diagram is evaluated in the same way with an exchange $x_2\leftrightarrow x_4$ which gives
     \begin{align}
        \mathcal{A}_t(1234) \, =\, { 2\normalization^2 \over |x_{12}|^2|x_{34}|^2} \sum_{s} c_s^2 u G_{s+1,s}({1\over u},{v\over u})\label{eq: s channel 4}
\end{align}
agreeing with the $t$-channel OPE expressions (for $2A+2C$).
Likewise, the $u$-channel contribution is
    \begin{align}
        \mathcal{A}_u(1234)={2\normalization^2 \over |x_{12}|^2|x_{34}|^2} \sum_{s} c_s^2 {u\over v} G_{s+1,s}(v,u)\label{eq: s channel 5}
\end{align}
which is consistent with the $u$-channel OPE expressions (for $2B+2C$).
The exchange $x_2\leftrightarrow x_3$ and $x_2\leftrightarrow x_4$ leads to a transformation of $(A,B,C)$ among each other
\begin{align}
    (x_2\;\leftrightarrow\; x_3)\quad :&\quad (A,B,C)\quad\longrightarrow \quad (A,C,B)\\
    (x_2\;\leftrightarrow\; x_4)\quad :&\quad (A,B,C)\quad\longrightarrow \quad (C,B,A)
\end{align}
and, by using this transformation, we have
\begin{equation}
    (A+B) + (A+B)_{2\leftrightarrow 3}+ (A+B)_{2\leftrightarrow 4}= 4(A+B+C)\, .\label{eq: transformation of ABC}
\end{equation}
Note that the $s$-channel contribution $\mathcal{A}_s(1234)$ in \eqref{eq: s channel 3} is $2(A+B)$. Hence, the sum of the $s$-, $t$- and $u$-channel contributions can be written as follows
\begin{equation}
    \mathcal{A}_s(1234)+\mathcal{A}_t(1234)+\mathcal{A}_u(1234) = 4 (A+B+C)\, .
\end{equation}

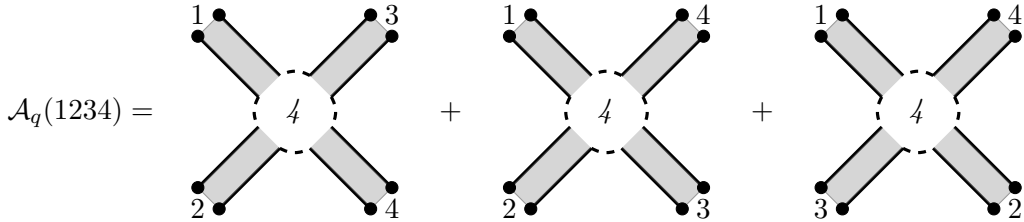
\begin{figure}[t!]
\centering
\begin{minipage}[c]{0.15\linewidth}
\centering
$\mathcal{A}_q(1234)=$
\end{minipage}
\begin{minipage}[c]{0.2\linewidth}
\centering
\begin{tikzpicture}
\draw [line width=1.2pt] (0.483,0.2)-- (1.29,1.007);
\draw [line width=1.2pt] (0.2,0.483)-- (1.007,1.29);
\draw [line width=1.2pt] (-0.483,0.2)-- (-1.29,1.007);
\draw [line width=1.2pt] (-0.2,0.483)-- (-1.007,1.29);
\draw [line width=1.2pt] (0.483,-0.2)-- (1.29,-1.007);
\draw [line width=1.2pt] (0.2,-0.483)-- (1.007,-1.29);
\draw [line width=1.2pt] (-0.483,-0.2)-- (-1.29,-1.007);
\draw [line width=1.2pt] (-0.2,-0.483)-- (-1.007,-1.29);
\draw (-1.29,1.29) node {1};
\draw (-1.29,-1.29) node {2};
\draw (1.29,1.29) node {3};
\draw (1.29,-1.29) node {4};
\draw [line width=1.2pt, dashed] (0.2,0.483)  to [bend angle = 30, bend right] (-0.2,0.483);
\draw [line width=1.2pt, dashed] (-0.2,-0.483)  to [bend angle = 30, bend right] (0.2,-0.483);
\draw [line width=1.2pt, dashed] (0.483,0.2)  to [bend angle = 30, bend left] (0.483,-0.2);
\draw [line width=1.2pt, dashed] (-0.483,0.2)  to [bend angle = 30, bend right] (-0.483,-0.2);
\draw [fill=black!40,opacity=0.4] (0.483,0.2)-- (1.29,1.007) -- (1.007,1.29) -- (0.2,0.483);
\draw [fill=black!40,opacity=0.4] (-0.483,0.2)-- (-1.29,1.007) -- (-1.007,1.29) -- (-0.2,0.483);
\draw [fill=black!40,opacity=0.4] (0.483,-0.2)-- (1.29,-1.007) -- (1.007,-1.29) -- (0.2,-0.483);
\draw [fill=black!40,opacity=0.4] (-0.483,-0.2)-- (-1.29,-1.007) -- (-1.007,-1.29) -- (-0.2,-0.483);
\draw [fill] (1.29,1.007) circle [radius=.08];
\draw [fill] (1.007,1.29) circle [radius=.08];
\draw [fill] (-1.29,1.007) circle [radius=.08];
\draw [fill] (-1.007,1.29) circle [radius=.08];
\draw [fill] (1.29,-1.007) circle [radius=.08];
\draw [fill] (1.007,-1.29) circle [radius=.08];
\draw [fill] (-1.29,-1.007) circle [radius=.08];
\draw [fill] (-1.007,-1.29) circle [radius=.08];
\draw (0,0) node {\textit{4}};
\end{tikzpicture}
\end{minipage}
\begin{minipage}[c]{0.05\linewidth}
\centering
$+$
\end{minipage}
\begin{minipage}[c]{0.2\linewidth}
\centering
\begin{tikzpicture}
\draw [line width=1.2pt] (0.483,0.2)-- (1.29,1.007);
\draw [line width=1.2pt] (0.2,0.483)-- (1.007,1.29);
\draw [line width=1.2pt] (-0.483,0.2)-- (-1.29,1.007);
\draw [line width=1.2pt] (-0.2,0.483)-- (-1.007,1.29);
\draw [line width=1.2pt] (0.483,-0.2)-- (1.29,-1.007);
\draw [line width=1.2pt] (0.2,-0.483)-- (1.007,-1.29);
\draw [line width=1.2pt] (-0.483,-0.2)-- (-1.29,-1.007);
\draw [line width=1.2pt] (-0.2,-0.483)-- (-1.007,-1.29);
\draw (-1.29,1.29) node {1};
\draw (-1.29,-1.29) node {2};
\draw (1.29,1.29) node {4};
\draw (1.29,-1.29) node {3};
\draw [line width=1.2pt, dashed] (0.2,0.483)  to [bend angle = 30, bend right] (-0.2,0.483);
\draw [line width=1.2pt, dashed] (-0.2,-0.483)  to [bend angle = 30, bend right] (0.2,-0.483);
\draw [line width=1.2pt, dashed] (0.483,0.2)  to [bend angle = 30, bend left] (0.483,-0.2);
\draw [line width=1.2pt, dashed] (-0.483,0.2)  to [bend angle = 30, bend right] (-0.483,-0.2);
\draw [fill=black!40,opacity=0.4] (0.483,0.2)-- (1.29,1.007) -- (1.007,1.29) -- (0.2,0.483);
\draw [fill=black!40,opacity=0.4] (-0.483,0.2)-- (-1.29,1.007) -- (-1.007,1.29) -- (-0.2,0.483);
\draw [fill=black!40,opacity=0.4] (0.483,-0.2)-- (1.29,-1.007) -- (1.007,-1.29) -- (0.2,-0.483);
\draw [fill=black!40,opacity=0.4] (-0.483,-0.2)-- (-1.29,-1.007) -- (-1.007,-1.29) -- (-0.2,-0.483);
\draw [fill] (1.29,1.007) circle [radius=.08];
\draw [fill] (1.007,1.29) circle [radius=.08];
\draw [fill] (-1.29,1.007) circle [radius=.08];
\draw [fill] (-1.007,1.29) circle [radius=.08];
\draw [fill] (1.29,-1.007) circle [radius=.08];
\draw [fill] (1.007,-1.29) circle [radius=.08];
\draw [fill] (-1.29,-1.007) circle [radius=.08];
\draw [fill] (-1.007,-1.29) circle [radius=.08];
\draw (0,0) node {\textit{4}};
\end{tikzpicture}
\end{minipage}
\begin{minipage}[c]{0.05\linewidth}
\centering
$+$
\end{minipage}
\begin{minipage}[c]{0.2\linewidth}
\centering
\begin{tikzpicture}
\draw [line width=1.2pt] (0.483,0.2)-- (1.29,1.007);
\draw [line width=1.2pt] (0.2,0.483)-- (1.007,1.29);
\draw [line width=1.2pt] (-0.483,0.2)-- (-1.29,1.007);
\draw [line width=1.2pt] (-0.2,0.483)-- (-1.007,1.29);
\draw [line width=1.2pt] (0.483,-0.2)-- (1.29,-1.007);
\draw [line width=1.2pt] (0.2,-0.483)-- (1.007,-1.29);
\draw [line width=1.2pt] (-0.483,-0.2)-- (-1.29,-1.007);
\draw [line width=1.2pt] (-0.2,-0.483)-- (-1.007,-1.29);
\draw (-1.29,1.29) node {1};
\draw (-1.29,-1.29) node {3};
\draw (1.29,1.29) node {4};
\draw (1.29,-1.29) node {2};
\draw [line width=1.2pt, dashed] (0.2,0.483)  to [bend angle = 30, bend right] (-0.2,0.483);
\draw [line width=1.2pt, dashed] (-0.2,-0.483)  to [bend angle = 30, bend right] (0.2,-0.483);
\draw [line width=1.2pt, dashed] (0.483,0.2)  to [bend angle = 30, bend left] (0.483,-0.2);
\draw [line width=1.2pt, dashed] (-0.483,0.2)  to [bend angle = 30, bend right] (-0.483,-0.2);
\draw [fill=black!40,opacity=0.4] (0.483,0.2)-- (1.29,1.007) -- (1.007,1.29) -- (0.2,0.483);
\draw [fill=black!40,opacity=0.4] (-0.483,0.2)-- (-1.29,1.007) -- (-1.007,1.29) -- (-0.2,0.483);
\draw [fill=black!40,opacity=0.4] (0.483,-0.2)-- (1.29,-1.007) -- (1.007,-1.29) -- (0.2,-0.483);
\draw [fill=black!40,opacity=0.4] (-0.483,-0.2)-- (-1.29,-1.007) -- (-1.007,-1.29) -- (-0.2,-0.483);
\draw [fill] (1.29,1.007) circle [radius=.08];
\draw [fill] (1.007,1.29) circle [radius=.08];
\draw [fill] (-1.29,1.007) circle [radius=.08];
\draw [fill] (-1.007,1.29) circle [radius=.08];
\draw [fill] (1.29,-1.007) circle [radius=.08];
\draw [fill] (1.007,-1.29) circle [radius=.08];
\draw [fill] (-1.29,-1.007) circle [radius=.08];
\draw [fill] (-1.007,-1.29) circle [radius=.08];
\draw (0,0) node {\textit{4}};
\end{tikzpicture}
\end{minipage}
\caption{Contact Diagrams}
\label{fig: contact diagrams}
\end{figure} 
To conclude our $s$, $t$, $u$ channel exchange evaluations, we note that the only contributions come from physical intermediate states (associated with single trace operators, and physical states of higher spin AdS fields).
For comparison in analogous evaluations of~\cite{Bekaert:2014cea}, there seem to appear contributions associated with additional double trace `states'~\cite{Bekaert:2016ezc}.

Next to complete the evaluation of the four point function of the scalar operator $\mathcal{O}$ in the free $O(N)$ vector model, we turn to evaluation of the quartic contact diagram contribution.
As in the exchange diagrams, one can simplify the contribution of the contact diagrams in Figure~\ref{fig: contact diagrams} to obtain
\begin{align}
    \mathcal{A}_q (1234) = -3(A+B+C)\, .\label{eq: contact diagram contribution}
\end{align}
%
%
Finally, by summing up the contributions of the exchange and contact diagrams, we have
\begin{align}
    \langle \mathcal{O}(x_1)\mathcal{O}(x_2)\mathcal{O}(x_3)\mathcal{O}(x_4) \rangle_{\text{\tiny conn}} &= \mathcal{A}_s(1234)+\mathcal{A}_t(1234)+\mathcal{A}_u(1234)+\mathcal{A}_q(1234)\cr
    &= (A+B+C)\, .
\end{align}

Let us summarize the main features of the above Feynman diagram scheme.
First, we comment on the feature that (for the free case) exchange and contact diagrams give essentially the same contribution. It is interesting to state the mechanism behind this since it relates to questions of non-locality. The vertices in the bi-local construction can be explicitely seen to be associated with derivative couplings and a bi-local star product. In the exchange diagrams (of the free theory ) we see that the derivatives in the 3-vertices cancel out the bi-local Laplacians of the propagator, effectively resulting in a contact type contribution. It is for this reason that the two type of contributions give similar results. In the bi-local basis neither the quartic, nor the higher contact vertices contain inverse Laplacian type terms. There is another aspect (of the free theory) that illuminates this issue.
It was observed that the Mellin amplitude in this case degenarates, instead of $1/s$, $1/t$, $/1/u$ type exchange terms one finds $x\delta(x)$ type terms which relates to the cancellation noted above. The amplitude is essentially 0, which again agrees with the bi-local S-matrix finding~\cite{deMelloKoch:2012vc}.

Concerning extensions of the diagram scheme to the IR fixed point it will be given by the same collective vertices, the difference being in the propagator, where the lowest mode wavefunction has to be taken to adjust to the different boundary conditions. Next, as we have also stressed the evaluation of loops is feasible, since the fact that the formalism is based on physical higher spin degrees of freedom avoids the need for ghost fields \etc. This was already seen in a previous evaluation of free energy~\cite{Jevicki:2014mfa}. It will be nevertheless interesting to compare the loop diagrams given by the present approach with the recent AdS evaluations of~\cite{Giombi:2017hpr}.

\section{Conclusion}
\label{sec:conclusion}

We have in this paper developed the details of the covariant bi-local holographic reconstruction of bulk AdS applicable to cases of vectorial AdS duality.
Different theories are distinguished by different bi-local propagators, while the all order interaction vertices are universally specified by the collective field action.
The following two key ingredients of the construction have been elaborated in detail.
First is the correspondence (Map) between the bi-local space and AdS space with higher spins.
This map is seen (as in our previous Hamiltonian constructions) to be one-to-one, with the bi-local space faithfully representing the physical content of the emergent bulk fields.
Even more (in the cases $d=1,2$) it was a purely momentum space Map, corresponding to a change of momentum space coordinates.
In this and other respects, the bi-local space reconstruction differs from schemes such as ``kinetic space'',
the later featuring maps~\cite{Czech:2016xec} from ($d+d$) dimensions to ($d+1$) representing a projection.
Moreover, a comparison of Laplacians between the two schemes would indicate that the later might not have a satisfactory $1/N$ basis.
In the bi-local construction, the reconstruction is fully nonlinear, with all order interactions which turn into bulk AdS space interactions through the one-to-one Map. As such they produce Feynman (Witten) diagrams applicable at tree and loop levels.
We have presented in detail the evaluations of 3 and 4-point cases, the latter resulting in the familiar conformal block expansion~\cite{Hijano:2015zsa}.
This in our $s$, $t$ and $u$ channel Feynman exchanges was seen to receive contributions only from single traces, representing physical states (in the example presented). For the example of free vector theory,a particular simplification is noted, where the derivative couplings in the vertices cancel out the Laplacians in the propagators, with a result being of contact type. It is not the case for the nontrivial IR fixed point theory, where one obtains genuine exchange and contact type contributions.

Finally, we mention that other theories which are not necessarily based on vector fields (such as matrix and tensor models) offer sectors governed by bi-local holography. This is the case also of the recent interesting example of `fishnet' theories~\cite{Grabner:2017pgm}, which can be seen to correspond to a simple deformation of the present case. A notable challenge is still de Sitter space-time, which might be reachable through bi-local holography. For a recent, interesting discussion see~\cite{Neiman:2018ufb}.

\acknowledgments

One of us (AJ) would like to acknowledge conversations with Sacha Zhiboedov which trigered some of the evaluations presented. We would also like to thank Sumit Das for his continuing interest and discussions on the present topic. The work of RdMK is supported by the South African Research Chairs Initiative of the Department of Science and Technology and National Research Foundation as well as funds received from the National Institute for Theoretical Physics (NITheP). The work of AJ and KS is supported by the Department of Energy under contract DE-SC0010010. The work of KS is also supported by the Galkin Fellowship Award at Brown University,
and received funding from the European Research Council (ERC) under the European Union's Horizon 2020 research and innovation programme under grant agreement No 758759. JY gratefully acknowledges support from International Centre for Theoretical Sciences (ICTS), Tata Institute of Fundamental Research, Bengaluru. RdMK and JY thank the Yukawa Institute for Theoretical Physics at Kyoto University, where this work was completed during the workshop YITP-T-18-04 ``New Frontiers in String Theory 2018''. JY also thanks Asia Pacific Center for Theoretical Physics~(APCTP) for the hospitality and partial support within the program ``Holography and Geometry of Quantum Entanglement''. KS and JY would like to thank the Nordic Institute for Theoretical Physics (Nordita) for the hospitality and partial support during the completion of this work, within the program ``Bounding Transport and Chaos in Condensed Matter and Holography''. 


\appendix

\section{Fourier Transform of Spinning Correlators}
\label{app:fourier transform of spinning correlators}

From~\cite{Costa:2011mg}, the 2-dimensional spinning conformal correlators are given by 
	\begin{equation}
		\cpw_{\Delta, s}(\vec{x}_1, \vec{x}_2; \vec{x}_3;\epsilon) \, \sim \, \frac{(Z_{\mu} \epsilon^{\mu})^s}{|x_3-x_1|^{\Delta} |x_3-x_2|^{\Delta} |x_1-x_2|^{-\Delta}} \, , 
	\end{equation}
with
\begin{equation}
	Z^\mu \equiv {|x_{13}||x_{23}|\over |x_{12}| }\left(  {x_{13}^\mu \over |x_{13}|^2}- {x_{23}^\mu \over |x_{23}|^2} \right)\, .
\end{equation}
In this Appendix, we consider the Fourier transform of these spinning conformal correlators in the Lorentzian signature.

The arbitrary polarization $\epsilon^\mu$ can be removed using the Thomas operator defined as follows
	\begin{equation}
		\thomas_\mu \, = \, \epsilon^\nu\partial_\nu \partial_\mu -{1\over 2} \epsilon_\mu \partial_\nu \partial^\nu \, .
	\end{equation}
The form of the Thomas operator ensures that the tensor that we obtain after the $\epsilon^\mu$ have been removed by differentiation, is traceless. As usual in $2d$-CFT, it will prove convenient to use complex coordinates. Expressed in terms of complex coordinates, the Thomas operator becomes
	\begin{equation}
		\thomas_+ \, = \, \epsilon^+ \, \partial^2_+ \quad , \qquad \thomas_- \, = \, \epsilon^- \, \partial^2_- \, ,
	\end{equation}
and $(\epsilon\cdot Z)$ for the two possible (holomorphic or anti-holomorphic) polarizations becomes
	\begin{equation}
		Z^+ \, = \, - {|x_{13}||x_{23}|\over |x_{12}| } \, {x_{12}^-\over  x_{23}^- x_{13}^-}  \quad , \qquad
		Z^- \, = \, - {|x_{13}||x_{23}|\over |x_{12}| } \, {x_{12}^+\over  x_{23}^+ x_{13}^+}  \, .
	\end{equation}
Therefore, we have
	\begin{equation}
		Z_\mu \epsilon^\mu \, = \, {1\over 2} {|x_{13}||x_{23}|\over |x_{12}| } \left( {\epsilon^+ x_{12}^-\over  x_{13}^- x_{23}^-} + {\epsilon^- x_{12}^+ \over  x_{13}^+ x_{23}^+} \right) \, .
	\end{equation}
Now, the Fourier transform of the spinning conformal correlators for the holomorphic polarization are written as
	\begin{align}
		\cpw_{\Delta, s}(\vec{x}_1, \vec{x}_2; \vec{p}) \, \propto \, |\eta^+ \eta^-|^{-\Delta_0} \, e^{-i \vec{p} \cdot \vec{x}} &\int dx_3^+ dx_3^- \,
		e^{-ip_+ x_3^+} e^{-i p_- x_3^-} \nonumber\\
		&\ \times \left( \frac{\eta^+}{(x_3^+)^2 - (\eta^+)^2 } \right)^{\frac{\Delta+s}{2}}
		\left( \frac{\eta^-}{(x_3^-)^2 - (\eta^-)^2 } \right)^{\frac{\Delta-s}{2}} \, , 
	\label{psi_hs}
	\end{align}
where we defined
	\begin{equation}
		\vec{x} \, \equiv \, \frac{\vec{x}_1+\vec{x}_2}{2} \quad , \qquad \vec{\eta} \, \equiv \, \frac{\vec{x}_1 - \vec{x}_2}{2} \, .
	\label{x&eta}
	\end{equation}
The integrals in~\eqref{psi_hs} are given by the integral representation of modified Bessel functions as
	\begin{equation}
		\cpw_{\Delta, s}(\vec{x}_1, \vec{x}_2; \vec{p}) \, \propto \,
		\frac{|p_+|^{\nu^+} |p_-|^{\nu^-}}{\Gamma(\nu^++\tfrac{1}{2}) \Gamma(\nu^-+\tfrac{1}{2})} \ 
		|\eta^+\eta^-|^{\frac{1}{2}-\Delta_0} \, e^{-i \vec{p} \cdot \vec{x}} \, K_{\nu^+}(i p_+ \eta^+) \, K_{\nu^-}(i p_- \eta^-)
	\end{equation}
where we defined 
	\begin{align}
		\nu^+ \, \equiv \, \frac{\Delta+s-1}{2} \, , \qquad \nu^- \, \equiv \, \frac{\Delta-s-1}{2} \, .
	\end{align}

We further Fourier transform from $\vec{x}_1, \vec{x}_2$ to $\vec{k}_1, \vec{k}_2$:
	\begin{equation}
		\cpw_{\Delta, s}(\vec{k}_1, \vec{k}_2; \vec{p}) \, \equiv \, \int d^2\vec{x}_1 d^2\vec{x}_2 \, e^{i \vec{k}_1 \cdot \vec{x}_1} e^{i \vec{k}_2 \cdot \vec{x}_2} \, \cpw_{\Delta, s}(\vec{x}_1, \vec{x}_2; \vec{p}) \, .
	\end{equation}
Changing the integration variables to $\vec{x}$ and $\vec{\eta}$ as in~\eqref{x&eta}, then the $\vec{x}$ integral leads to the momentum conservation delta function $\delta^2(\vec{k}_1+\vec{k}_2-\vec{p})$.
The remaining $\vec{\eta}$ integrals can be directly performed by for example using formula 6.699.4 of \cite{Gradshteyn:1994} as
	\begin{align}
		&\quad \, \int_{-\infty}^{\infty} d\eta \, |\eta|^{\mu} \, e^{i k \eta} \, K_{\nu}(i p \eta) \nonumber\\
		&= \, 2^{\mu} (ip)^{-\mu-1} \, \Gamma\Big(\frac{\mu+\nu+1}{2}\Big) \Gamma\Big(\frac{\mu+\nu+1}{2}\Big) \,
		{}_2F_1\Big(\frac{\mu+\nu+1}{2}, \frac{1+\mu-\nu}{2}, \frac{1}{2}; \frac{k^2}{p^2}\Big) \, .
	\end{align}

There is a one possibility that this result becomes simpler. Namely when $\mu=0$ ($\Delta_0=1/2$), we have
	\begin{equation}
		\int_{-\infty}^{\infty} d\eta \, e^{i k \eta} \, K_{\nu}(i p \eta) \, = \, \frac{1}{\sqrt{1-\frac{k^2}{p^2}}} \, \cos\Big[ \nu \arcsin(\frac{k}{p}) \Big] \, .
	\end{equation}
Hence for $\Delta_0=1/2$, we obtain the Fourier transform of the spinning correlators as
	\begin{align}
		\cpw_{\Delta, s}(\vec{k}_1, \vec{k}_2; \vec{p})
		\, &\propto \, \frac{\delta^2(\vec{k}_1+\vec{k}_2-\vec{p})}{\Gamma(\frac{h+s}{2}) \Gamma(\frac{h-s}{2})} \,
		\frac{|p_+|^{\nu^+} |p_-|^{\nu^-}}{((p_+)^2 - (k_+)^2)^{\frac{1}{2}} ((p_-)^2 - (k_-)^2)^{\frac{1}{2}}} \nonumber\\
		&\qquad \times \cos\left[ \nu^+ \arcsin\left(\frac{k_+}{p_+} \right) \right]
		\cos\left[ \nu^- \arcsin\left(\frac{k_-}{p_-} \right) \right] \, .
	\label{3pt in momentum}
	\end{align}

\section{Checks for the Bi-local Map}
\label{app:checks for the bi-local map}

\subsection{Realization of $SO(2,2)$ Generators}
\label{app: realization of so(2,2)}

\begin{align}
		L_+^{\rm AdS} \, &= \, - ip_+ \, , \nonumber\\
		L_0^{\rm AdS} \, &= \, i\Big(p_+ x^+ + {1\over 2}p^z z\Big) \, , \nonumber\\
		L_-^{\rm AdS} \, &= \, -i\Big(p_+(x^+)^2 +  x^+ z p^z +  z^2 p_- + { (p^\theta)^2\over 4p_+} + {\sqrt{4p_+ p_- - (p^z)^2}\over 2 p_+}  z p^\theta \Big) \, , \nonumber\\
		\bar{L}_+^{\rm AdS} \, &= \, - ip_- \, ,\nonumber\\
		\bar{L}_0^{\rm AdS} \, &= \, i\Big(p_- x^- + {1\over 2}p^z z \Big)\, , \nonumber\\
		\bar{L}_-^{\rm AdS} \, &= \, -i \Big(p_- (x^-)^2 + x^- z p^z +  z^2 p_+ +  { (p^\theta)^2\over 4p_-}   - {\sqrt{4p_+ p_- - (p^z)^2}\over 2 p_- } z p^\theta \Big) \label{eq: lorentzian ads gens} \, ,
\end{align}

\subsection{Coordinate Map}
\label{app:coordinate map}

Using the inverse momentum map in \eqref{eq: inv momentum map1} and \eqref{eq: inv momentum map2}, we have
\begin{align}
    x \, =& \, {k_1 u_1 +k_2 u_2\over p}-{\sqrt{k_1}\sqrt{\bar{k}_1}-\sqrt{k_2}\sqrt{\bar{k}_2} \over 2p}\xi\label{eq: coordinate map1}\\
    \bar{x} \, =& \, {\bar{k}_1 \bar{u}_1 + \bar{k}_2 \bar{u}_2\over \bar{p}}-{\sqrt{k_1}\sqrt{\bar{k}_1}-\sqrt{k_2}\sqrt{\bar{k}_2} \over 2\bar{p}}\xi \label{eq: coordinate map2}\\
    z \, =& \,-{i\over 2} \xi \label{eq: coordinate map3}\\
    p^\theta \, =& \, -\sqrt{k_1} \sqrt{k_2}(u_1 -u_2) +  \sqrt{\bar{k}_1}\sqrt{\bar{k}_2 }(\bar{u}_2 -\bar{u}_1)\label{eq: coordinate map4}
\end{align}
where $\xi$ is defined by
\begin{equation}
    \xi\equiv {u_1-u_2\over {\sqrt{k_1}\over \sqrt{\bar{k}_1}}+ {\sqrt{k_2}\over \sqrt{\bar{k}_2}} }+ {\bar{u}_1-\bar{u}_2\over {\sqrt{\bar{k}_1}\over \sqrt{k_1}}+ {\sqrt{\bar{k}_2}\over \sqrt{k_2}} }
\end{equation}

\subsection{Transformation of Casimir and Laplacian}
\label{app:laplacian transformation}

For AdS$_3$ $\times$ $S^1$, we use momentum-space coordinates ($p, \bar{p}, \phi, \theta$).
The mapping to bi-local momentum-space is given by 
	\begin{alignat}{3}
		p \, &=& \, k_1 \, + \, k_2 \, , \qquad 	\phi \, =& \,   {1\over 2}\arcsin \frac{\kdif}{p} \, + {1\over 2} \arcsin \frac{\kdifb}{\bar{p}} \, , \nonumber\\
		\bar{p} \, &=& \, \bar{k}_1 \, + \, \bar{k}_2
	 \, , \qquad \theta \, =& \,{1\over 2} \arcsin \frac{\kdif}{p} \, - \, {1\over 2} \arcsin \frac{\kdifb}{\bar{p}} \, , 
	\label{bi-local map app}
	\end{alignat}
where we defined $k \equiv k_1 - k_2$, $k_a \equiv \frac{1}{2}(k_a^0 -i k_a^1)$, $p \equiv \frac{1}{2}(p^0 - i p^1)$, while
	\begin{equation}
		\phi \, \equiv \, \sinh^{-1}\left( \frac{p_z}{2\sqrt{p \bar{p}}} \right) \, .
	\end{equation}
We have
	\begin{equation}
		L_{\rm AdS_3 \times S^1}^2 \, = \, \frac{1}{4} \left( \frac{\partial}{\partial \phi} \, + \, \frac{\partial}{\partial \theta} \right)^2 \, , \qquad
		\overline{L}_{\rm AdS_3 \times S^1}^2 \, = \, \frac{1}{4} \left(  \frac{\partial}{\partial \phi} \, - \, \frac{\partial}{\partial \theta} \right)^2 \, .
	\end{equation}
Therefore, the Laplacian $\Box_{\rm AdS_3 \times S^1}^2 = 4 L^2 \overline{L}^2$ is given by
	\begin{equation}
		\Box_{\rm AdS_3 \times S^1} \, = \, \frac{1}{2} \left( \frac{\partial^2}{\partial \phi^2} \, - \, \frac{\partial^2}{\partial \theta^2} \right) \, .
	\end{equation}

For bi-local CFT, we label momentum-space by ($k_1, \bar{k}_1, k_2, \bar{k}_2$).
The inverse map of~\eqref{bi-local map app} is given by 
	\begin{gather}
		k_1 \, = \, \frac{p}{2} \Big[ 1 + \sin(\phi + \theta) \Big] \quad , \qquad
		\bar{k}_1 \, = \, \frac{\bar{p}}{2} \Big[ 1 + \sin(\phi - \theta) \Big] \, , \nonumber\\
		k_2 \, = \, \frac{p}{2} \Big[ 1 - \sin(\phi + \theta) \Big] \quad , \qquad
		\bar{k}_2 \, = \, \frac{\bar{p}}{2} \Big[ 1 - \sin(\phi - \theta) \Big] \, .
	\end{gather}
We have 
	\begin{equation}
		L_{\rm bi}^2 \, = \,  \left[ \sqrt{p^2 - \kdif^2} \ \frac{\partial}{\partial \kdif} \right]^2 \, , \qquad
		\overline{L}_{\rm bi}^2 \, = \, \left[ \sqrt{\bar{p}^2 - \kdifb^2} \ \frac{\partial}{\partial \kdifb} \right]^2 \, .
	\end{equation}
From the mapping~\eqref{bi-local map app}, one can rewrite the derivatives with respect to $k,\bar{k}$ as derivatives of $\phi$ and $\theta$, by employing the chain rule.
Therefore, we obtain
	\begin{equation}
		L_{\rm bi}^2 \, \to \, \frac{1}{4} \left( \frac{\partial}{\partial \phi} \, + \, \frac{\partial}{\partial \theta} \right)^2 \, = \, L_{\rm AdS_3 \times S^1}^2 \, , \qquad
		\overline{L}_{\rm bi}^2 \, \to \, \frac{1}{4} \left( \frac{\partial}{\partial \theta} \, - \, \frac{\partial}{\partial \phi} \right)^2 \, = \, \overline{L}_{\rm AdS_3 \times S^1}^2 \, .
	\end{equation}
This guarantees that the Laplacian is also correctly transformed from the bi-local CFT to AdS$_3$ $\times$ $S^1$.

\section{Conformal Block Expansions}
\label{app:spectrum form of the bi-local propagator}

The four point correlation function of spin zero scalars is given by
\begin{align}
    &\langle \mathcal{O}(x_1)\mathcal{O}(x_2)\mathcal{O}(x_3)\mathcal{O}(x_4) \rangle_{\text{\tiny conn}}
    \,=\, A+B+C\cr
    =&\,\frac{\Upsilon_{\mathcal{O}}^2}{(y_{12}^2 y_{34}^2)^{2\Delta_0}} \, \left(\sum_{s\ge0, \, {\rm even}} c_s^2 \, G_{s+d-2, s}(u, v)+\sum_s c_{n,s}^2 G_{s+2n+2,s}(u,v)\right)\cr
\end{align}
%
$c_s$ is the OPE coefficient for single trace operators
%
%
\begin{equation}
	c_s^2\,\equiv\, {1\over  N}  {2^{s+3} [({1\over 2})_s]^2 \over s! s_s}\label{eq: ope single trace}
\end{equation}
and $c_{n,s}$ is the OPE coefficient of double-trace operator~\cite{Dolan:2000ut,Diaz:2006nm} given by
\begin{equation}
    c_{n,s}^2\,\equiv\, {1\over N} {(-1)^n 4 ({1\over 2})_{n+{s\over 2} } [(s+n)! ({1\over 2})_{n}]^2 \over n! (n+{s\over 2})! [({s\over 2}!)]^2 (s+{3\over 2})_n (n)_n (s+2n+1)_s (s+n+{1\over 2})_n } \, .
\end{equation}
The contributions from each of the three channels is 
\begin{align}
        \mathcal{A}_s(1234)\, =&\,
        \frac{{8\over N}\Upsilon_{\mathcal{O}}^2}{(x_{12}^2 x_{34}^2)^{2\Delta_0}} \, \left[ \, u^{\Delta_0}\, + \, \left( \frac{u}{v} \right)^{\Delta_0}  \, \right] \, ={2 \normalization^2 \over |x_{12}|^2|x_{34}|^2} \sum_{s} c_s^2 G_{s+1,s}(u,v)\, ,\label{eq: s channel 3}\\
        \mathcal{A}_t(1234)\, =&\, \frac{{8\over N} \Upsilon_{\mathcal{O}}^2}{(x_{12}^2 x_{34}^2)^{2\Delta_0}} \, \left[ \, u^{\Delta_0} \, + \, \left( \frac{u^2}{v} \right)^{\Delta_0} \, \right]\, ={2 \normalization^2 \over |x_{12}|^2|x_{34}|^2} \sum_{s} c_s^2 u G_{s+1,s}({1\over u},{v\over u})\, ,\label{eq: t channel 3}\\
        \mathcal{A}_u(1234)\, =&\, 
        \frac{{8\over N}\Upsilon_{\mathcal{O}}^2}{(x_{12}^2 x_{34}^2)^{2\Delta_0}} \, \left[ \, \left( \frac{u}{v} \right)^{\Delta_0} \, + \,\left( \frac{u^2}{v} \right)^{\Delta_0}  \, \right] 
         \, ={2 \normalization^2 \over |x_{12}|^2|x_{34}|^2} \sum_{s} c_s^2 {u\over v}G_{s+1,s}(v,u)\, .\label{eq: u channel 3}
\end{align}

\section{Conformal Partial Wave Expansion}
\label{app: app title}

\subsection{CPW Basis}
\label{app: conformal partial wave}

Let us consider the conformal partial wave functions constructed from two free $O(N)$ scalar fields of dimension $\Delta_0\equiv {d-2\over 2}\equiv{d\over 2}+c_0$ and spin-$s$ symmetric-traceless operator of conformal dimension $\Delta={d\over 2}+\cdim$, which up to normalization are given by
\begin{equation}
	\cpw^{\mu_1\cdots \mu_s}_{\cdim,s} (x_1,\cdim_0;x_2,\cdim_0;x_3,\cdim_0,s)\sim {1\over N} \sum_{i=1}^N\langle\phi_i(x_1)\phi_i(x_2)\mathcal{O}_{\Delta={d\over 2} + \cdim }^{\mu_1\cdots \mu_s}(x_3)\rangle
\end{equation}
One can package the conformal partial wave function with a null polarization vector $\varepsilon^\mu$
\begin{align}
	\cpw_{\cdim,s} (x_1,\cdim_0;x_2,\cdim_0;x_3,\cdim,s,\varepsilon)&\equiv  \varepsilon_{\mu_1} \varepsilon_{\mu_2}\cdots  \varepsilon_{\mu_s}\cpw_{\cdim}^{\mu_1\cdots \mu_s} (x_1,\cdim_0;x_2,\cdim_0;x_3,\cdim,s) \cr
	\sim& { (Z_{12,3}\cdot  \varepsilon )^s\over |x_{12}|^{2\Delta_0 - \Delta} |x_{23}|^{\Delta} |x_{31}|^\Delta }
\end{align}
where $Z^\mu_{12,3}$ is a unit vector defined by
\begin{equation}
	Z^\mu_{12,3} \equiv {|x_{13}||x_{23}|\over |x_{12}| }\left(  {x_{13}^\mu \over |x_{13}|^2}- {x_{23}^\mu \over |x_{23}|^2} \right)\hspace{8mm} (\; Z^\mu_{12,3}Z_{12,3 \mu}=1\;)\, .
\end{equation}
In~\cite{Dobrev:1976vr,Dobrev:1977qv}, the conformal partial wave function $\cpw_{c,s}(x_1,x_2;x_3)$
\begin{align}
	\cpw_{\cdim,s}(x_1,\cdim_0;x_2,\cdim_0;x_3,\cdim,s,\varepsilon)\equiv & {1\over (2\pi)^{d\over 2} } { 2^{{3d\over 2} + \cdim +s +2 \cdim_0\over 2} \norm_{\cdim_0,s}( \cdim )\over |x_{12}|^{2\Delta_0 -\Delta}|x_{23}|^{\Delta}|x_{31}|^{\Delta} } (Z_{12,3}\cdot \varepsilon )^s\cr
	&(\Delta\equiv  {d\over 2}+ \cdim\;\;,\;\; \Delta_0\equiv  {d\over 2}+ \cdim_0)
\end{align}
has been studied, and its normalization constant $\norm(\cdim)$ was determined to be
\begin{align}
	\norm_{\cdim_0,s}(\cdim) \equiv&  \left[{\Gamma({d\over 4}+{\cdim \over 2} +{s\over 2}+c_0 )\Gamma({d\over 4}-{\cdim \over 2} +{s\over 2}+c_0  )[\Gamma({d\over 4}+{\cdim \over 2} +{s\over 2} )]^2\over\Gamma({d\over 4}+{\cdim \over 2} +{s\over 2} -c_0 )\Gamma({d\over 4}-{\cdim \over 2} +{s\over 2} -c_0 )[\Gamma({d\over 4}-{\cdim \over 2} +{s\over 2} )]^2 }\right]^{1\over 2}\, .\label{eq: normalization}
\end{align}
Note that for the free $O(N)$ scalar field we have $c_0\equiv -1$. 
%
%
The tensor product of two principal series of representations can be decomposed into principal series representations in any dimension, and for fixed pure imaginary $\tilde{c}$ (principal series) the conformal partial wave functions $\cpw(x_1,\tilde{\cdim};x_2,\tilde{\cdim};x_3,\cdim)$ with $c\in i\mathbb{R}$ form a complete basis~\cite{Dobrev:1976vr,Dobrev:1977qv}. For our purpose we need an analytic continuation of this completeness relation from pure imaginary $\tilde{c}$ to the real value $c_0=-1$ for the free $O(N)$ scalar field. As $\tilde{\cdim}$ is analytically continued to $\cdim_0=-1$, poles from the normalization factor $\norm_{\cdim_0,s}(\cdim)$ might cross the contour $\cdim\in (-i\infty, i\infty)$ in the completeness relation\footnote{This divergence comes from 
\begin{equation}
	\norm_{\tilde{\cdim},s }(-\cdim)N_{\tilde{\cdim} ,s}(\cdim)={\Gamma({d\over 4}+{\cdim \over 2} +{s\over 2}+\tilde{\cdim} )\Gamma({d\over 4}-{\cdim \over 2} +{s\over 2}+\tilde{\cdim} ) \over\Gamma({d\over 4}+{\cdim \over 2} +{s\over 2} -\tilde{\cdim} )\Gamma({d\over 4}-{\cdim \over 2} +{s\over 2} -\tilde{\cdim} ) }
\end{equation}
} provided that~\cite{Dobrev:1977qv}
\begin{equation}
	{d\over 2} +s+2\cdim_0  < 0
\end{equation}  
For $s\geqq 2$, the contour integral does not pick up any pole in the analytic continuation. However, for $s=0$ the contour can cross a pole. In particular, for $d=3$ we have the extra contribution of $\Delta=1,2$ which corresponds to the pole at $\cdim=\pm {1\over 2}$. This leads to the completeness relation for the conformal partial wave constructed from the $O(N)$ free scalar field
\begin{align}
	&\sum_{\cdim,s}  \int d^dy {\lambda_{\cdim,s} \rho_s(\cdim) \over  |x_{34}|^{4}} \;  \cpw_{\cdim,s} (x_1,\cdim_0;x_2,\cdim_0;y,- \cdim ,s,\partial_\epsilon) \cpw_{\cdim,s} (x_3,\cdim_0;x_4,\cdim_0;y, \cdim ,s,\epsilon)\cr
	=&{1\over 2} \delta^d(x_{13})\delta^d(x_{24})+{1\over 2} \delta^d(x_{14})\delta^d(x_{23}) \label{eq: cpw completeness}
\end{align}
where the Plancherel weight $\rho_s(\cdim)$ is defined by
\begin{equation}
	\rho_s(\cdim)\equiv {\Gamma({d\over 2} +s)\over 2 (2\pi)^{d\over 2}s!}\left|{\Gamma(d/2-1+\cdim)\over \Gamma(\cdim)}\right|^2 [(d/2+s-1)^2-\cdim^2]
\end{equation}
First of all, note that one can think of the summation over $\cdim$ and $s$ in \eqref{eq: cpw completeness} as the vertical contour integral (in $\cdim$ space) along the pure imaginary axis together with two residues at $\cdim=\pm {1\over 2}$ for $s=0$, or equivalently as the deformed contour $\mathcal{C}'$ to include the pole at $c=-{1\over 2}$ and to exclude the other pole at $c={1\over 2} $ only for the $s=0$ case
\begin{align}
	\sum_{s,\nu}\equiv  \left[\int_{-i \infty}^{i\infty} {d\cdim \over 2\pi i}+\res_{\cdim={1\over 2}}-\res_{\cdim=-{1\over 2}}\right]_{s=0}+\sum_{s=1}^\infty \int_{-i \infty}^{i\infty} {d\cdim \over 2\pi i}=\left[\int_{\mathcal{C}' } {d\cdim \over 2\pi i }\right]_{s=0}+\sum_{s=1}^\infty \int_{-i \infty}^{i\infty}  {d\cdim \over 2\pi i}\label{eq: def summation}
\end{align}
The deformed contour $\mathcal{C}'$ is a common prescription for the extra contribution coming from the analytic continuation~\cite{Simmons-Duffin:2017nub}. But, as pointed out in \cite{Simmons-Duffin:2017nub}, one can also think of these contributions as the contribution of finite numbers of ``discrete'' representations for $s=0$ which are called ``Born terms'' in the literature\cite{Mack:2009mi}.

Without these ``discrete'' contributions, the contour integral for $s=0$ in the $s$-channel would pick up a simple pole at $\cdim ={1\over 2}$ which corresponds to the massless representation $D(2,0)$ of the scalar field in the $AdS_4$. The extra contribution will eliminate this pole and add a simple pole which was located outside of the original contour. This additional simple pole corresponds to the massless representation $D(1,0)$ of the scalar field in $AdS_4$.

Also, note that the factor ${\lambda_{c,s} \over  |x_{12}|^4}$ in \eqref{eq: cpw completeness} comes from transforming one of the conformal partial wave functions in the completeness relation into its shadow
\begin{align}
	\cpw_{\cdim,s} (x_1,1;x_2,1 ;y,\cdim,s,\varepsilon)={\lambda_{c,s} \over  |x_{12}|^4} \cpw_{\cdim,s} (x_1, -1;x_2,-1 ;y,\cdim,s,\varepsilon)
\end{align}
where $\lambda_{\cdim,s}$ is the eigenvalue of the bi-local Laplacian~in \eqref{def: bilocal laplacian} given by
\begin{equation}
	\lambda_{\cdim,s}=
	{1\over 4}\left[ \cdim^2 -(s+{d\over 2}-2)^2\right]\left[ \cdim^2 -(s+{d\over 2})^2 \right]\, .
\end{equation}
%

%
%
%
%
%
%
%
%
The bi-local fluctuation $\eta(x_1,x_2)$ can be expanded in terms of the conformal partial wave function
\begin{equation}
	\eta(x_1,x_2) = \sum_s \int {d\cdim \over 2\pi i}\int d^d y \;\eta_{\cdim, s}(y,\partial_\epsilon) \cpw_{\cdim,s}(x_1,\cdim_0;x_2,\cdim_0;y,\cdim,s,\epsilon)
\end{equation}
where $\int {d\cdim \over 2\pi i}$ is either the deformed contour or vertical contour along $(-i\infty,i\infty)$ (in $\cdim $ space) with two small circles around two poles. Note that the conformal partial wave function is symmetric or anti-symmetric for even or odd spins, respectively \ie
\begin{equation}
	\cpw_{\cdim,s}(x_2,\cdim_0;x_1,\cdim_0;y,\cdim,s,\epsilon) =(-1)^s \cpw_{\cdim,s}(x_1,\cdim_0;x_2,\cdim_0;y,\cdim,s,\epsilon)
\end{equation}
Since the bi-local field for the $O(N)$ vector model is symmetric, only conformal partial wave functions with even spin appear.

\subsection{Identity}
\label{app: identity}

We can write the conformal block as~\cite{Kos:2014bka}
\begin{equation}
	G_{\Delta,s} (u,v)= \mathfrak{G}_s(\Delta;u,v) +\sum_i {a_i\over \Delta-\Delta_i }G_{\Delta_i +n_i, s_i}(u,v) 
\end{equation}
where $\mathfrak{G}_s(\Delta;u,v)$ is an analytic function in complex $\Delta$ space. In \cite{Kos:2014bka}, the poles that may appear are classified into three posible classes. Since our contour integral for the $s$-channel encloses the region $\text{Re} (\cdim)>0$, it is enough for us to consider the third class of poles. These have the form
\begin{align}
	G_{{3\over 2}+\cdim,s}\supset \sum_{k=1}^s {\hat{a}_3(k,s)\over \cdim - ({1\over 2} +s-k)}G_{2+s,s-k}(u,v)\label{eq: expansion}
\end{align}
where $\hat{a}_3(k,s)$ is given\footnote{Note that we follow the convention of~\cite{Penedones:2015aga} for conformal blocks, and it is different from that in~\cite{Kos:2014bka} (See~\eqref{eq: convention of blocks}). Hence, the coefficient $\hat{a}_3(k,s)$ is modified according to the differences in conventions.} by
\begin{equation}
	\hat{a}_3(k,s)= -2^{-k}{({d-2\over 2})_{s-k} \over ({d-2\over 2})_s}{ (d-2)_s\over (d-2)_{s-k} }{ k \over (k!)^2}{(s+1-k)_k [({1-k\over 2})_k]^2\over (s+{d\over 2}-k)_k}\, .
\end{equation}
From now on we specialize to $d=3$. First of all, for each $s$, the pole at $\cdim=s-{1\over 2}$ (\ie $k=1$) corresponds to a single-trace operator. The corresponding residue vanishes (\ie $\hat{a}_3(1)=0$) so that the conformal block $G_{{3\over 2}+\cdim,s}(u,v)$ does not have a pole at $\cdim=s-{1\over 2}$.

For a given $s$, from \eqref{eq: expansion} we see that the conformal block $G_{{3\over 2}+i\nu,s+k}$ contains the following simple pole
\begin{align}
	G_{{3\over 2}+\cdim,s+k}\supset {\hat{a}_3(k,s+k)\over \cdim - ({1\over 2} +s)}G_{s+k+2,s}(u,v)\hspace{8mm} (k=2,3,\cdots )\, . 
\end{align}
Note that this pole at $\cdim=s+{1\over 2}$ of $G_{{3\over 2}+\cdim,s+k}$ $(k=2,3,\cdots)$ is distinct from the poles of ${\rho_s(\cdim)  \over  \lambda_{\cdim,s}}K_{\cdim,s}$ in \eqref{eq: s channel 2}. Hence, the $s$-channel contribution in~\eqref{eq: s channel 2} picks up a simple pole at $\cdim = s+{1\over 2}$ to give the conformal block $G_{s+k+2,s}(u,v)$ as a residue. For $k=2n+1$, one can see that
\begin{equation}
	\hat{a}_3(2n+1,s+2n+1)=0 \hspace{7mm}(n=1,2,\cdots)\, .
\end{equation}
Therefore, the simple pole of the conformal block for $k=2n+2$ $(n=0,1,2,\cdots)$ gives a non-zero contribution to the $s$-channel contribution in~\eqref{eq: s channel 2}. The corresponding residue is proportional to $G_{s+2n+4,s}(u,v)$ \ie
\begin{align}
	G_{{3\over 2}+\cdim,s+2n+2}\supset {\hat{a}_3(2n+2,s+2n+2)\over \cdim- ({1\over 2} +s)}G_{s+2n+4,s}(u,v)\hspace{8mm} (n=0,1,2,\cdots)\, . \label{eq: pole of conformal block}
\end{align}
The contribution of these poles exactly cancels with the contribution from a second type also contributing to the $s$-channel amplitude, namely, the simple poles at $\cdim =s+2n+{5\over 2}$ ($n=0,1,2,\cdots$ and $s=0,2,4,\cdots$) from the ${\rho_s(\cdim)  \over  \lambda_{\cdim,s}}K_{\cdim,s}$. To see this, one can show that 
\begin{align}
	&\sum_{s} \res_{\cdim=s+2n+{5\over 2} }  {2^{{5\over 2}+s}\over (2\pi)^3}{\rho_s(\cdim)  \over  \lambda_{\cdim,s}}K_{\cdim,s}G_{{3\over 2}+\cdim,s}(u,v)\cr
	&+\sum_{s} \res_{\cdim=s+{1\over 2}}  {2^{{5\over 2}+s+2n+2}\over (2\pi)^3}{\rho_{s+2n+2}(\cdim)  \over  \lambda_{\cdim,s+2n+2}}K_{\cdim,s+2n+2}G_{{3\over 2}+\cdim,s+2n+2}(u,v)=0\label{eq: identity to prove}
\end{align}
By explicit evaluation, one can easily verify that the coefficients of the conformal block $G_{s+2n+4,s}(u,v)$ in \eqref{eq: identity to prove} vanish
\begin{align}
	\left[\left(\res_{\cdim=s+2n+{5\over 2} }  {\rho_s(\cdim)  \over  \lambda_{\cdim,s}}K_{\cdim,s}\right)+  2^{2n+2}  {\rho_{s+2n+2}(s+{1\over 2} )  \over  \lambda_{s+{1\over 2},s+2n+2}}K_{s+{1\over 2},s}\hat{a}_3(2n+2,k+2n+2)\right]=0\ ,
\end{align}
which proves \eqref{eq: identity to prove}.


\bibliographystyle{JHEP}
\bibliography{covariant}

\end{document}